\documentclass[%
 aip,
 jmp,%
 amsmath,amssymb,
 reprint,%
]{revtex4-1}

\usepackage[T1]{fontenc}
\usepackage[latin9]{inputenc}
\usepackage{float}
\usepackage{textcomp}
\usepackage{amsmath}
\PassOptionsToPackage{normalem}{ulem}

\usepackage{ulem}
\usepackage{color}
\usepackage{graphicx}
\usepackage{dcolumn}
\usepackage{bm}
\usepackage[mathlines]{lineno}
\usepackage{url}
\usepackage{hyperref}

\usepackage{epsfig}
\usepackage{epstopdf}
\usepackage{amsthm}
\graphicspath{{.}{./FIGS/}{./FIGS2/}}

\definecolor{OliveGreen}{rgb}{0,0.6,0}
\definecolor{applegreen}{rgb}{0.55,0.71,0.0}
\definecolor{darkpastelgreen}{rgb}{0.01,0.75,0.24}

\begin{document}

\preprint{AIP/123-QED}

\title{Small-angle neutron scattering and Molecular Dynamics structural
study of gelling DNA nanostars}

\author{{J. Fernandez-Castanon}}
 \affiliation{Sapienza--Università di Roma, P.le A. Moro 5, 00185 Roma, Italy}
 
 \author{F. Bomboi}
\affiliation{Sapienza--Università di Roma, P.le A. Moro 5, 00185 Roma, Italy}

\author{L. Rovigatti}
\affiliation{\mbox{Rudolf Peierls C. T. P., University of Oxford, 1 Keble Road, Oxford, OX1 3NP, United Kingdom} }
\affiliation{\mbox{Faculty of Physics, University of Vienna, Boltzmanngasse 5, A-1090 Vienna, Austria}}

\author{M. Zanatta}
\affiliation{\mbox{Dipartimento di Fisica, Università di Perugia, Via Pascoli, 06123 \emph{}Perugia, Italy }}

\author{A. Paciaroni}
\affiliation{\mbox{Dipartimento di Fisica, Università di Perugia, Via Pascoli, 06123 \emph{}Perugia, Italy }}

\author{L. Comez}
\affiliation{\mbox{Dipartimento di Fisica, Università di Perugia, Via Pascoli, 06123 \emph{}Perugia, Italy }}
\affiliation{IOM-CNR, UOS Perugia c/o Dipartimento di Fisica e Geologia, Università di Perugia, \\Via Pascoli, 06123 Perugia, Italy}

\author{L. Porcar}
\affiliation{\mbox{Institut Laue-Langevin, 71 avenue des Martyrs, CS 20156, 38042 Grenoble cedex 9, France}}

\author{C. J. Jafta}
\affiliation{\mbox{Helmholtz-Zentrum Berlin, Hahn-Meitner-Platz 1, 14109 Berlin, Germany}}

\author{G. C. Fadda}
\affiliation{\mbox{Laboratoire Léon Brillouin, LLB, CEA Saclay, 91191 Gif-sur-Yvette Cedex, France}}

\author{T. Bellini}
\affiliation{Department of Medical Biotechnology and Translational Medicine, Università di Milano, \\I-20133 Milano, Italy}

\author{F. Sciortino }
 \email{francesco.sciortino@phys.uniroma1.it}
\affiliation{Sapienza--Università di Roma, P.le A. Moro 5, 00185 Roma, Italy}
\affiliation{CNR-ISC, UOS Sapienza--Università di Roma, I-00186 Roma, Italy}

\date{\today}

\begin{abstract}
DNA oligomers with properly designed sequences self-assemble into
well defined constructs. Here, we exploit this methodology to produce
bulk quantities of tetravalent DNA nanostars (each one composed by 196
nucleotides) and to explore the structural signatures of their aggregation
process. We report small-angle neutron scattering experiments focused
on the evaluation of both the form factor and the temperature evolution
of the scattered intensity at a nano star concentration where the
system forms a tetravalent equilibrium gel. We also perform molecular
dynamics simulations of one isolated tetramer to evaluate the form factor
theoretically, without resorting to any approximate shape. The numerical
form factor is found to be in very good agreement with the experimental
one. Simulations predict an essentially temperature independent form
factor, offering the possibility to extract the effective structure
factor and its evolution during the equilibrium gelation. 
%
\end{abstract}

\maketitle

\section{\label{sec:Introduction}Introduction}

The development of new \emph{smart nanomaterials} \cite{2008__APP_SmartMaterials,2006_APP_MolecularNanotech},
which are able to adapt their response to different external stimuli
at the nanoscale, paves the way for the future of fields as diverse
as medicine \cite{2008_Impact_Nanotechn_medicine_dentistry,2001_APP_ContactLenses,2006_APP_Medicine_Pepas},
 drug-delivery \cite{2003_APP_Kopecek_DrugDel,2014_APP_CancerTherapy_Portugal,2005_APP_DrugDel_Tachibana},
photonics \cite{2007_APP_Photonics_NewMaterial,2011_APP_Photonics_DNA} and
 computing \cite{1996_APP_Computing_Turing1,2012_APP_Computing_nanorobots},
among others. The desoxyribonucleic acid (DNA) is one of the most
promising materials to encompass all of the aforementioned applications
due to its base pairing specificity (A···T, G···C) which allows for
absolute control over the design of deliberated structures. The responsible of storing our genetic information, something as natural
as life itself, startlingly provides the perfect ingredient to create new functional
materials \cite{2003_Seeman_DNA_Material_World,2009_APP_MaterialsScienceDNA}
via a cascade of self-assembly processes, each one guided by the length
of complementary sequences of distinct DNA strands.

\smallskip{}

One of the first DNA constructs which has been designed and realized
in the lab~\cite{zadegan2012structural} is the DNA nanostar (NS)
with controlled valence $f$. To build the structure, $f$ properly
designed DNA single strands are mixed together in equimolar concentrations. 
 
\begin{figure}[t]
\hspace*{-0.5cm}  
\raggedright{}\includegraphics[scale=0.25]{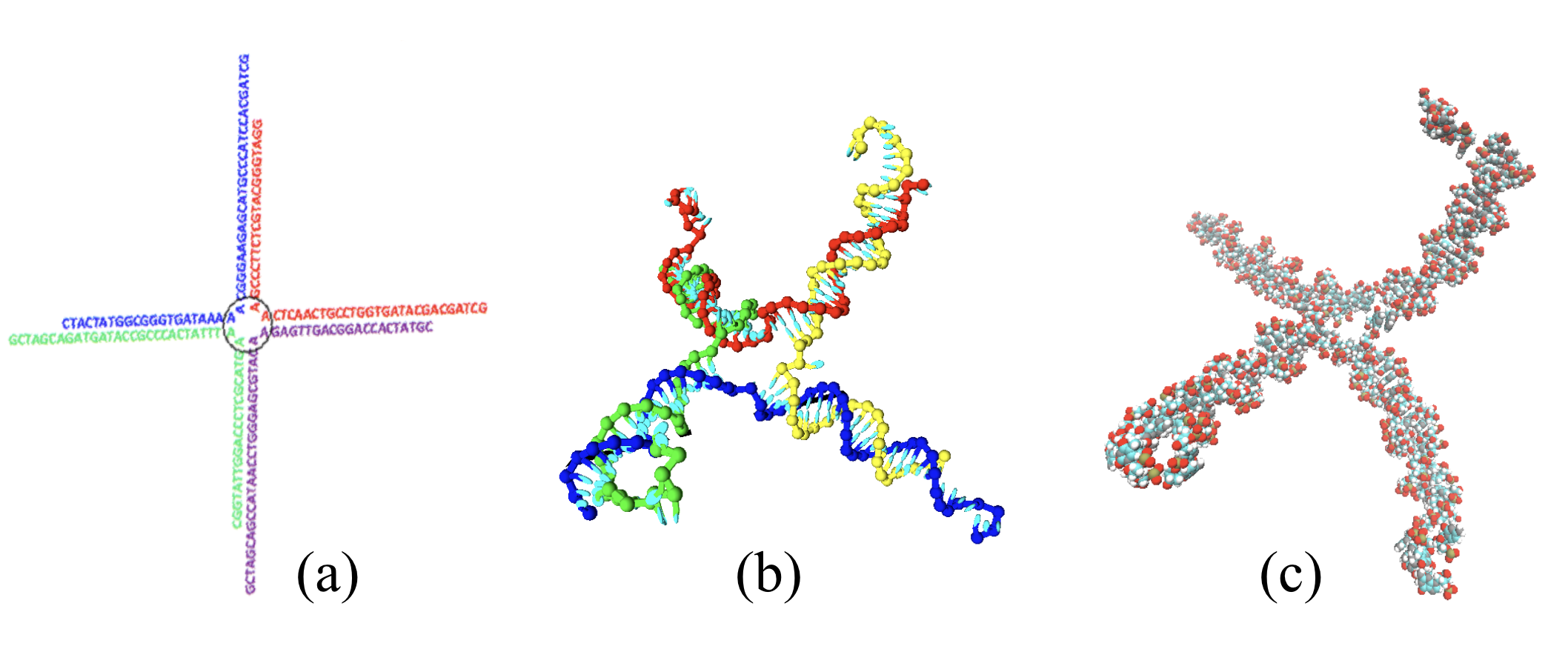}
\caption{\label{fig:immagine} Representation of a NS at different levels.
(a) The four sequences of bases forming the NS. Each single strand
has been represented with a different color. Note the six unpaired
bases, acting as sticky ends (b) the oxDNA representation of the self-assembled
NS, in which each base is modelled as a rigid body; (c) the corresponding
full atom representation.}
\end{figure}

The sequence design favours the pairing of each
strand with two different partners, in such a way that a 
construct with a flexible, unpaired core and $f$
double-helical arms structure is spontaneously
formed (see Fig.~\ref{fig:immagine}). A self-complementary short sequence
at the end of each arm provides the sticky site to bind distinct
NSs. This methodology allows for the synthesis of supramolecular constructs
in bulk quantities, opening the way for an experimental study of their
bulk behavior. NS particles have been selected as optimal candidates
for testing the role of the valence on the gas-liquid phase separation
\cite{biffi2013phase,2015_Equilibrium_gels_DNA_nanostars_Bomboi}.
Consistent with theoretical studies~\cite{bianchi2006phase,2008_Zaccarelli_Sciortino}
it has been shown that these particles undergo a phase-separation
process between a phase of isolated NS and a network phase, in which
NS particles bind to form a thermoreversible gel, the physical analog
of $f$-functional chemical gels~\cite{2008_Sakai_Shibayama}.

Differently from the chemical analog, the equilibrium phase behavior
of this system can be explored. Beyond the coexisting density, on
cooling, the system moves continuously from a high-$T$ state in which
monomers only interact via their excluded volume (and eventually electrostatic)
repulsion to a fully bonded low temperature state (the equilibrium
gel) via a progressive formation of larger and larger aggregates.
The possibility to break and reform bonds makes it possible to release
stresses and reach at low $T$ a fully bonded state, e.g. a gel free
of entanglement and defects.

Dynamic light scattering (DLS) experiments~\cite{2015_Equilibrium_gels_DNA_nanostars_Bomboi,bomboi2015equilibrium}
of the DNA-gel formation have shown that the density fluctuations
relax slower and slower on cooling, following an Arrhenius $T$ dependence,
increasing by more than five orders of magnitude in a small $T$ interval,
before exiting from the accessible experimental time window (10 seconds).
Here, we present a series of Small-Angle Neutron Scattering (SANS)
experiments carried out to quantify the structural signatures associated
with the formation of the equilibrium gel at a suitably selected concentration
as a function of $T$. At this concentration, the system evolves
from fluid to gel   through a succession of equilibrium
steps, without the interference of phase separation. We also present
experiments at a much lower concentration, where the form factor of
the NS can be measured. We complement the experimental study with
a numerical investigation of the structure of a single NS, based on
the coarse-grained oxDNA2~\cite{doye2013coarse} model. This provides an effective way to
connect the observed signatures in the SANS diffraction pattern with
geometrical parameters. The quality of the experimental data and the
agreement with the theoretical evaluation of the form factor allow
us to extract an effective structure factor between the centers of
the NSs under the  hypothesis of decoupling
between translational and orientational degrees of freedom.

\smallskip{}

\section{Materials and Methods}

\subsection*{Materials and Sample preparation}

The four sequences programmed to self-assemble in tetravalent DNA-NSs 
are:

\smallskip{}

\textbf{Sequence 1. }$5'$-{\color{darkpastelgreen} CTACTATGGCGGGTGATAAA}\textcolor{red}{AA}\\
 {\color{blue}CGGGAAGAGCATGCCCATCC}\textcolor{red}{A}\textbf{CGATCG}-$3'$

\textbf{Sequence 2.} $5'$-{\color{blue}GGATGGGCATGCTCTTCCCG}\textcolor{red}{AA}\\
 {\color{cyan}CTCAACTGCCTGGTGATACG}\textcolor{red}{A}\textbf{CGATCG}-$3'$

\textbf{Sequence 3.} $5'$-{\color{cyan}CGTATCACCAGGCAGTTGAG}\textcolor{red}{AA}\\
 {\color{magenta}CATGCGAGGGTCCAATACCG}\textcolor{red}{A}\textbf{CGATCG}-$3'$

\textbf{Sequence 4.} $5'$-{\color{magenta}CGGTATTGGACCCTCGCATG}\textcolor{red}{AA}\\
{\color{darkpastelgreen}TTTATCACCCGCCATAGTAG}\textcolor{red}{A}\textbf{CGATCG}-$3'$

\smallskip{}

where sequences with the same color indicate complementary strands.
The red \textcolor{red}{AA} bases at the center of each sequence constitute
the central flexible core. The black (\textbf{CGATCG}) self-complementary
$6$-bases overhangs provide the sequences that permit the connection
between different DNA-NS via so-called \emph{sticky-ends.} Bonding
between different DNA-NS is favoured by the inclusion of a nucleotide
\textcolor{red}{A} immediately before the sticky-end. The non-bonded
sequences help releasing angular and rotational constraints, permitting
both arm bending and rotation of the end sequences. Depending on the
salt-concentration, NSs form at relatively high $T$ and the lifetime
of the aggregates becomes essentially infinite at ambient and smaller
$T$. In our case, NSs self-assemble around 65\textdegree C and start
to bind to each other below $50$\textdegree C~\cite{biffi2013phase}.

We prepare the sample dissolving each of the four DNA single strands
(provided by IDT and purified in a polyacrylamide gel (PAGE)) in deionized
and degassed filtered ($0.2\text{ }\mu\text{m}$ filters) H$_{2}$O
water. Each sample is then centrifuged at $25\text{ \textdegree C}$
/ $4.4$ rpm for $5$ minutes to favour the powder dissolution. Up
to three different Nanodrop~\cite{2010_Nanodrop} measurements are
undertaken to obtain accurate values of the strand concentration.
The absorbance measurements (ratios $260/280=1.89$ and $260/230=2.27$)
confirm the absence of proteins and low concentrations of other contaminants.
The resulting solutions are then mingled at  proper mixing ratios in order to obtain
an equimolar solution of the four different strands. The prepared final
concentration was $21$ mg/ml, corresponding to $348$ $\mu$M of
NS. Assuming that each phosphate group releases one counter ion, then
$[\text{Na}^{+}]\approx63$ mM\textcolor{black}{.}\textcolor{green}{{}
}The experimental form factor of the DNA-NS required the preparation
of a more diluted sample at $c=3.2\text{ mg ml}^{-1}$ ($53\text{ }\mu$M
DNA-NS). The diluted sample was prepared in an electrolyte $63\text{ mM}$
NaCl solution to mimic the Na release conditions of the high concentrated
sample. This DNA concentration meets the requirements to provide a
signal strong enough, and consequently a good statistics in the SANS
measurements, but at the same time to reduce to the minimum the inter-particle
interactions. The resulting samples are heated at $90$\textdegree C
for $20$ minutes and then cooled down to room temperature in about
$7$ hours. Further details on the preparation can be found in Ref.~\onlinecite{2015_Equilibrium_gels_DNA_nanostars_Bomboi}.

\subsection*{Small-Angle Neutron Scattering experiments}

Comparable volumes, \textcolor{black}{$80\text{ }\mu\text{l}$, of
the two samples (low and high concentrations) were prepared }to fill up the quartz Hellma cells
($0.5$ mm path). 

SANS measurements of the high concentrated sample were performed at
the small-angle diffractometer $D22$ of the Institut Laue Langevin
{{[}DOI: 10.5291/ILL-DATA.9-13-559{]}} (ILL, Grenoble, France)
\cite{2008_D33_ILL,http_www_ill_fr_d22}. Measurements were done at different temperature, ranging 
 from $55$\textdegree C to $5$\textdegree C. Immediately before the data
acquisition, we let the samples thermalize for $25$ minutes at each
temperature. Exploiting the ILL's high flux reactor, measurements
of $5$, $20$ and $30$ minutes were respectively taken at three
sample-detector distances $L_{SD}=1.40$ m, $L_{SD}=5.60$ m and $L_{SD}=17.00$
m. This instrument configuration, together with the incident neutron
wavelength $\lambda_{0}=6\text{}\textrm{ \AA}$ selected, allowed
us to cover a wavevector, $q$, window ranging from $0.0025$ to $0.6\text{}\textrm{ \AA}^{-1}$.

\smallskip{}

The raw data were treated according to standard procedures, including
solvent and empty cell subtraction, using GRASP software provided
by ILL \cite{2003__GRASP_Manual}, which yields the value of the scattering
intensity onto the absolute intensity scale.

The same sample was also probed at the SANS instrument PACE of the Laboratoire
Léon Brillouin (LLB, CEA Saclay, France).  The instrument configuration allowed to cover
a $q$-range from $0.02$ to $0.6\text{}\textrm{ \AA}^{-1}$. This was obtained with three setups,
using an incident neutron wavelength $\lambda_{0}=5\text{}\textrm{ \AA}$ combined to 
$L_{SD}=1.00$ m and $L_{SD}=3.00$ m, and with $\lambda_{0}=12\text{}\textrm{ \AA}$ and 
$L_{SD}=3.00$ m to reach the smallest $q$-values. In order to get a good signal-to-noise ratio, an 
acquisition time of $30$ minutes was necessary for the first two cases, extending it up to $60$ minutes for the last one.
To test the reversibility of the assembling process of our system, three measurements were taken at $T=20$\textdegree C.
Starting from $20$\textdegree C, the sample was heated up to $45$\textdegree C, cooled down back to $20$\textdegree C, then to $6$\textdegree C,
and finally heated up again to $20$\textdegree C. Each measurement was performed after a thermalization time of
$20$ minutes. The data were analysed with the software PASiNET.

Finally, for the purpose of properly evaluate the form factor $P(q)$, the more diluted sample was measured at the V4 instrument
of the neutron source BER II at the Helmoltz-Zentrum Berlin facilities
(HZB, Berlin, Germany) \cite{V4_BerII_Berlin}. After $40$ minutes
of sample thermalization at $50$\textdegree C, rather long measurements
of about $3$ hours, about $7$ hours, and $19$ hours were acquired 
at $L_{SD}=1.35$ m, $L_{SD}=4.00$ m and $L_{SD}=16.00$ m, respectively.
These sample-detector distances, coupled to an incident wavelength 
$\lambda_{0}=6\text{}\textrm{ \AA}$, 
allow to cover a $q$-range of $0.01<q<0.39\textrm{ \AA}^{-1}$,
equivalent to the one investigated on D22.
Standard data reduction was accomplished by means of the BerSANS-PC
software \cite{2002_BerSANS_PC}.

Deuteration, either of the solvent or of the sample, is a typical
procedure when dealing with biological samples. Typically, the exchange of
hydrogen with deuterium atoms permits to 
exploit the advantages of contrast variation in SANS
experiments, and thus to clearly discriminate between the buffer and the sample.
However, in this case, no contrast investigations were required. In fact,
the little amount of H atoms found in DNA when compared to proteins
and lipids, allowed us to use H$_{2}$O as a solvent to achieve the
best contrast condition, while removing most of the strong $^{1}$H
incoherent background via a careful buffer subtraction, retaining 
a  good statistics for the scattered signal.

\section{Simulations}

We perform molecular dynamics simulations of one isolated NS for different
$T$ and salt concentrations to provide theoretical predictions
for the NS form factor. We employ oxDNA2~\cite{snodin2015introducing},
a coarse-grained model that has been shown to provide a physical representation
of the thermodynamic and mechanical properties of single- and double-stranded
DNA~\cite{ouldridge2011structural,vsulc2012sequence,doye2013coarse}.
The basic unit of the model is a nucleotide, represented as an oriented
rigid body. In oxDNA2, the interaction between nucleotides takes into
account the sugar-phosphate backbone connectivity, excluded volume,
hydrogen bonding, nearest neighbour stacking, cross-stacking between
base-pair steps in a duplex and coaxial stacking. Hydrogen bonding
can occur between complementary bases when they are antialigned,
leading to the formation of double-stranded conformations. Electrostatic
interactions are included as screened Yukawa repulsions, assuming
dissociation of the phosphorous sites.

In order to evaluate the form factor of the NS, we first need to convert
the oxDNA2 representation into a full-atom one. This conversion is
crucial to properly reproduce the interatomic distances and therefore
their scattering signature in the large $q$ window. We carry out this
procedure by considering that the orientation of each coarse-grained
base is identified by three axes. Two of these, $\vec{a}_{1}$ and
$\vec{a}_{3}$, define the directions along which hydrogen bonding
and stacking interactions are maximised. These can be mapped onto
an aromatic base by exploiting the planarity of the latter, making
it possible to define the atomic analogues of $\vec{a}_{1}$ and $\vec{a}_{3}$.
This procedure fixes the orientation of the bases. Their positions 
are then set by superimposing the base site of each coarse-grained
nucleotide with the centre of mass of the full-atom aromatic ring
and shifting it by $1.13$ Å. When applied to a perfect double helix,
this method reproduces the full-atom phosphate-phosphate distances
with a $99.9$\% accuracy.

We evaluate the numerical form factor $P({\bf q})$ as the modulus

\begin{equation}
P({\bf q})\equiv<|F_{NS}({\bf q})|^{2}>
\end{equation}
where 
\begin{equation}
F_{NS}({\bf q})\equiv\sum_{j=1}^{N}b_{i}\exp[i{\bf q}\cdot{\bf t}_{j}],\label{eq:fq}
\end{equation}
$N$ is the total number of atoms composing the NS, ${\bf t}_{j}$
is the vector joining the center of mass of the NS to atom $j$ and
$b_{i}$ the atom scattering length~\cite{nist}. The average $<...>$
is performed over all different orientations and all different configurations
generated in the molecular dynamics run (to sample all possible geometrical
shapes assumed by the NS). We also evaluate 
\begin{equation}
F({\bf q})\equiv<F_{NS}({\bf q})>^{2}\label{eq:FormFactor_F(q)}
\end{equation}
a quantity requested to estimate the quality of the orientational
decoupling~\cite{chen2015scattering,kotlarchyk1983analysis}.

\section{Results and discussion}

\subsection{Dilute sample: Form Factor analysis}

\begin{figure}[t]
\hspace*{-0.5cm}  
\begin{centering}
\includegraphics[scale=0.36]{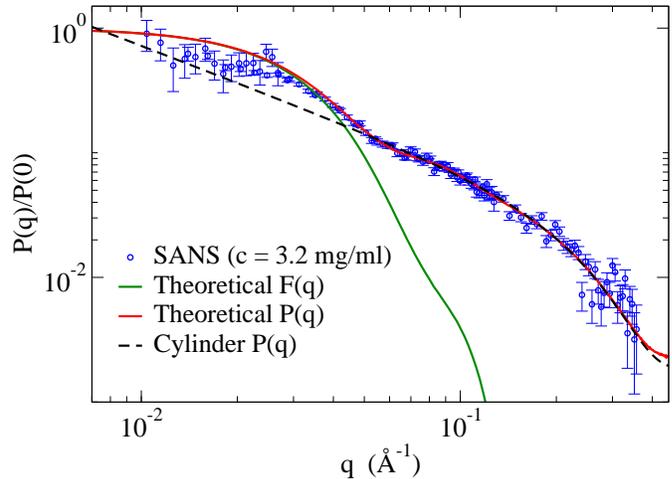} 
\par\end{centering}

\caption{\label{fig:form} Normalized form factor $P(q)$. Experimental SANS
(blue dots) and simulated (solid red line) form factor at $50$\textdegree C
as a function of $q$ in log-log scale. $P(q)$ of an infinite long
cylinder with cross section radius $R_{c}=8\textrm{ \AA}$ from Eq.~\ref{eq:cilindro}
(dashed black line). The figure shows also $F(q)$ from Eq.~\ref{eq:FormFactor_F(q)}
(solid green line). The numerical $P(q)$ and $F(q)$ are calculated
averaging over an equilibrium ensemble of configurations of an isolated
NS at $50$\textdegree C and {[}NaCl{]} =0.1 M. }
\end{figure}

\noindent Fig.~\ref{fig:form} shows the normalized intensity measured
at $T=50$\textdegree C for the sample at $c=3.2$ mg/ml, compared
with the form factor $P(q)$ calculated numerically from the NS generated
via the oxDNA2 potential~\cite{snodin2015introducing}, with full-atom 
substitution. Beyond 0.03 $\text{}\textrm{ \AA}^{-1}$, the experimental
data are very well described by the theoretical function, supporting
the quality of the oxDNA force-field in modelling the structure of
the NS. The gyration radius, calculated by the atomic model, correspond
to $R_{g}=54\text{}\textrm{ \AA}$, to be compared with an estimated
length of the double helix arms of 68 $\textrm{ \AA}$ (3.4 $\text{}\textrm{ \AA}$
for 10 bases~\cite{jones2002soft}). Fig.~\ref{fig:form} also shows
the theoretical form factor of a homogeneous rigid cylinder with length
much larger than the diameter ($2r_{c}$)~\cite{kassapidou1997structure,1955_Guinier_Cylinder_Pq}
\begin{equation}
P(q)\sim\left(\frac{2J_{1}(qr_{c})}{qr_{c}}\right)^{2}\label{eq:cilindro}
\end{equation}
where $J_{1}$ is the Bessel function of the first
kind. Between $0.1\textrm{ \AA}^{-1}<q<0.35\textrm{ \AA}^{-1}$ the signal from the
cylindrical shape of the DNA arms is prominent. Indeed, in this $q$-vector
window, the experimental data can be quite accurately modeled by the
form factor of a homogeneous rigid cylinder of radius $8\textrm{ \AA}$.
This cross-section radius agrees with the value reported in previous
small-angle scattering studies of DNA short double helices~\cite{1998_Short_DNA_fragments,1999_Maarel_small_Rc,1986_Short_DNA_fragment_SANS}.
We ascribe the difference between such a value and the outer diameter of the B-DNA helix ($\sim10\textrm{ \AA}$) to the assumption of homogeneous cylinder
\cite{1953_WatsonCrick_DNA_structure,1998_SANS_DNA_Polyelectrolyte,1976_Luzzati_10A_DNA,1976_GarciaTorre_10A_DNA,1995_Koch_10A_DNA}.

For $q<0.03\text{}\textrm{ \AA}^{-1}$, the experimental form factor
shows deviations from the simulated one, signalling the presence
of a repulsive interaction between the NS, possibly originated by
the screened electrostatic repulsion. Indeed, DNA is known to be a
highly negatively charged polymer, in which all phosphorous groups
are ionised, resulting in a bare net charge of about 200 $e$ per tetramer. 
The significant DNA charge originates an inter-NS repulsion significantly
larger than the thermal energy, resulting to a first approximation
in an expanded double helix diameter~\cite{1998_Short_DNA_fragments}.
A precise characterization of this low $q$ window would require measurements
at lower NS concentrations, where, unfortunately, experiments are not easily
performed due to the weak scattering signal.

The quality of the comparison between the experimental and theoretical
structure factor, beside providing a precise characterisation of the
geometry of the NS, confirms the high efficiency of the self-assembly
process. Electrophoretic gel runs have indeed suggested that more
than 95 per cent of the NS properly form~\cite{biffi2013phase}.

\begin{figure}[b]
\vspace*{-0.55cm}
\hspace*{-0.7cm} 
\noindent \begin{centering}
\includegraphics[scale=0.36]{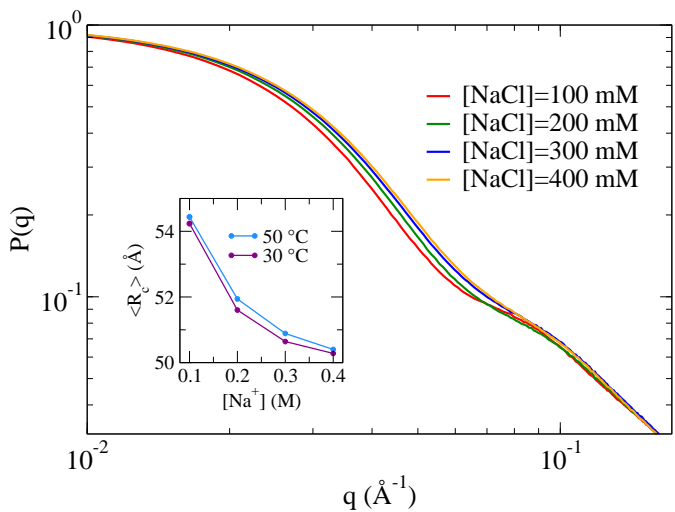} 
\par\end{centering}

\caption{\label{fig:Simul:Rg_T_Salt} Theoretical predictions for the ionic
strength dependence of the form factor based on the oxDNA2 model.
The inset shows the corresponding gyration radii $R_{g}$ for two
different $T$ as a function of salt concentration. }
\end{figure}

The agreement between simulations and experimental data offers us
the possibility to exploit simulations to quantify the dependence
of the NS radius of gyration, $R_{g}$ on $T$ and on the salt concentration.
The $T$-dependence is particularly relevant, since it is not possible
to perform experiments at low $T$ at dilute concentration due to
the onset of the limited-valence gas-liquid like phase separation~\cite{biffi2013phase}
which takes place in the sample. Luckily, the inset of Fig.~\ref{fig:Simul:Rg_T_Salt}
reveals that $R_{g}$ strongly depends on the salt concentration whereas
it weakly changes with $T$. According to the oxDNA2 model, the increase
of $R_{g}$ arises prevalently from the expansion of the central
junction, where repulsive forces are non compensated by complementary
base bonding. The predicted salt dependence of $P(q)$ shows the sensitivity
of the experimental measurement, which is only consistent with the form
factor evaluated at the lowest accessible ionic strength~\cite{snodin2015introducing}.

\subsection{Gel-forming sample}

Previous static and dynamic light scattering studies have focused on the thermodynamic
behavior of the NS, providing evidence of a limited-valence phase
separation~\cite{biffi2013phase} in the low concentration region.
At a total $[\text{Na}^{+}]\approx 60$ mM, phase separation extends
from very low concentration up to 17 mg/ml. For higher NS concentration
the system remains homogeneous at all temperatures, forming an open equilibrium-gel structure at low
$T$ ~\cite{biffi2013phase}.

We provide here the first measurement of the structural properties
of the system in the equilibrium gel region, covering the range of
$T$ (from $55$\textdegree C to $5$\textdegree C) over which DLS
 observes an increase of the relaxation time by more
than five orders of magnitude, revealing the formation of larger and
larger clusters of bonded NSs that eventually span the entire system,
forming the gel. The highest investigated $T$ is lower than the temperature
at which the stars unfold ($T_{m}\approx65$\textdegree C).

\begin{figure}[h]
\hspace*{-0.7cm}  
\noindent \begin{centering}
\includegraphics[scale=0.33, angle=-90]{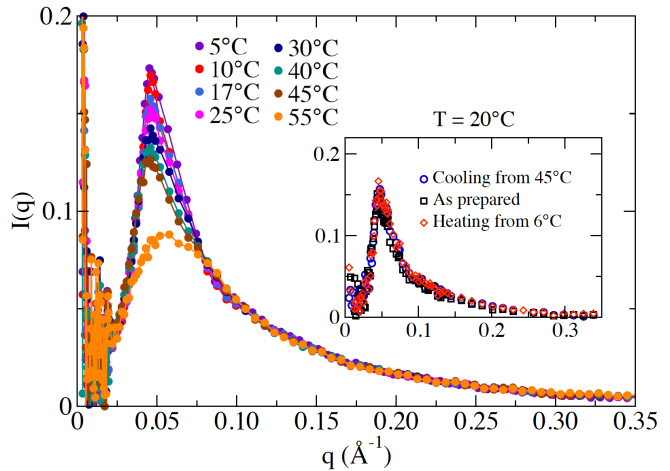} 
\par\end{centering}

\caption{SANS: Sample $c=21$ mg ml$^{-1}$. $I(q)$ in the $q$-range from
$0.0025<q<0.35$ $\textrm{ \AA}^{-1}$  (normalized to coincide with the normalized form factor at large $q$)
at temperatures varying from
$55$\textdegree C to $5$\textdegree C, measured at the D$22$ diffractometer. The inset (same units) shows three different measurements 
done at the PACE diffractometer, all performed at $20$\textdegree C
to provide evidence of full reversibility.  
The sample was initially measured at $20$\textdegree C after a long equilibration 
(black squares). Subsequently the sample was re-measured at  $20$\textdegree C  during a 
cooling scan started 
 from $45$\textdegree C
(blue circles) and finally  re-measured at  $20$\textdegree C  during a heating scan starting from $6$\textdegree C (red diamonds).}
\label{fig:SANS:Sample-Iq_vs_q} 
\end{figure}

The scattered intensity provides a measure of the space pair correlation,
weighted by the atomic scattering length. Formally, for a system of
$N_{S}$ nanostars, the $q$-dependence of the signal, is defined as~\cite{hansen1990theory,chen2015scattering}

\begin{equation}
I({\bf q})\equiv\frac{1}{N_{S}}<\sum_{l=1}^{N_{S}}\sum_{m=1}^{N_{S}}\exp[i{\bf q}\cdot({\bf r}_{_{CM}}^{l}-{\bf r}_{_{CM}}^{m})]F_{l}({\bf q})F_{m}^{*}({\bf q})>
\label{eq:iq}
\end{equation}

where ${\bf r}_{_{CM}}^{l}$ indicates the center of mass position
of the $l$-th NS and $F_{l}({\bf q})$ is the previously defined $F_{NS}$
function (see Eq.~\ref{eq:fq}) for the $l$-th NS.

Experimentally, the differential cross-section $d\sigma$ per solid angle $d\Omega$ is defined as
\begin{equation}
\left(\frac{d\sigma}{d\Omega}\right)_{NS}\kern-1.4em ({\bf q})=\beta I({\bf q})\label{eq:exp_xs}
\end{equation}
where ${\beta}=nv_{p}^{2}(\Delta \rho)^{2}$, with $n$ the number density of particles, $v_{p}$ the 
volume occupied by a NS, and $\Delta \rho$ the particle scattering length density difference between NS and solvent.

Since we have measured the solvent contribution only at one temperature, we have estimated the scattered
intensity from the system of DNA NSs, $(\frac{d\sigma_{c}}{d\Omega})_{NS}({q})$, at the different $T$ by subtracting
from the measured intensity, $I_{m}=(\frac{d\sigma_{c}}{d\Omega})_{m}({q})$, the solvent scattering, $I_{s}=(\frac{d\sigma_{c}}{d\Omega})_{s}({q})$,
multiplied by a fitting $T$-dependenet factor $\alpha$. 
The best value for  $\alpha$ has been determined by imposing that in the  $0.1<q<0.3\text{}\textrm{ \AA}^{-1}$ region
 the signal from NSs coincides (again apart from the constant $\beta$) with the normalized theoretical form factor $P(q)$, by defining
  
 \begin{equation}
 \left(\frac{d\sigma_{c}}{d\Omega}\right)_{NS}\kern-1.4em ({q}) = I_{m}(q)- \alpha I_{s}(q)
 \end{equation}
  and  finding $\alpha$ and $\beta $  by minimizing  the variance $\chi^2$
 $$
 \chi^2 \equiv \int_{q=0.1}^{q=0.3} \left [  \left(\frac{d\sigma_{c}}{d\Omega}\right)_{NS}\kern-1.4em ({q}) -  \beta P(q) \right ]^2 dq
 $$ 
The best fit values for $\alpha$  are all between $0.99$ and $0.96$, suggesting a very small temperature variation of the solvent scattering.  The best fit values of
$\beta$ (defining the value of the form factor in $q=0$) are found to be  $1.4 \le \beta \le 1.8$.

Figure~\ref{fig:SANS:Sample-Iq_vs_q} shows the resulting $I(q)$ (normalized by $\beta$)  at all investigated  
temperatures.
According to these results, we can differentiate three scattering
regions. At low-$q$ values, at the lower limit of the experimental
resolution, we observe a significant signal, suggesting the presence
of correlated scatterers over tens of nanometers. This upturn for
$q<0.02\textrm{ \AA}^{-1}$, which is commonly found in polyelectrolyte
solutions \cite{1995_Xanthan_upturn,2015_Polymer_upturn_India}, has
been widely discussed by several authors \cite{1996_Upturn_contr_sedlak,1998_SANS_DNA_Polyelectrolyte,1998_Shibayama_upturn_inhomog,1991_Matsuoka_upturn_discussion,2001_upturn_contr_zhang,2002_upturn_contr_sedlak}.
 Within this context, the strong low-$q$ signal is usually associated
to the recurrent clustering behavior of biological macromolecules\cite{1969_Clustering_Dusek,1982_cluster_Gelssier,1993_Geissler_Horkay_Cluster,1989_Clustering_Guenet}.
In the present case, this tendency is always observed, at all $T$,
even when all NS are not bounded and hence this very low $q$ peak
can not be associated with inhomogeneities in the gel. We note on
passing that DLS has also evidenced the presence
of small concentrations of approx 0.1 $\mu m$ size, which are possibly
introduced in the sample during the DNA synthesis. These impurities
could be well responsible for this low $q$ signal.

The most interesting part of the scattered intensity is the region
$q\approx0.05\textrm{ \AA}^{-1}$, where we observe the presence
of a peak which increases its amplitude on cooling. The corresponding
real-space distance, estimated as $d^{*}=2\pi/q=114\textrm{ \AA}$
is comparable with the center-to-center distance in a pair of bonded
NSs (estimated in about 140 $\textrm{ \AA}$), suggesting the possibility
to interpret such growth as structural evidence of the progressive
formation of the gel, in agreement with the previous DLS measurements 
of the same DNA NS sample solutions~\cite{2015_Equilibrium_gels_DNA_nanostars_Bomboi}.

For $q>0.1\textrm{ \AA}^{-1}$ the intensity does not vary with temperature
anymore and all the curves decay following the form factor.

It is worth noting that the aggregation process of the DNA NSs is fully reversible. This 
is clearly visible in the inset of Fig.~\ref{fig:SANS:Sample-Iq_vs_q}, where three
measurements at $T=20$\textdegree C are reported. The first measurement was 
taken at the sample preparation conditions, whereas the second and third were
acquired, always at $20$\textdegree C, after cooling the sample down from $T=45$\textdegree C and after heating it up from
$T=6$\textdegree C, respectively.

\subsection{Structure Factor Analysis}

Assuming the possibility
of decoupling the center of mass from the orientational variables,
it is customary to approximate $I(q)$ (Eq.~\ref{eq:iq}) as

\begin{eqnarray}
I({\bf q})=P({\bf q})+\hspace{3.5cm}\nonumber\\
\frac{1}{N_{S}}<\sum_{l=1}^{NS}\sum_{\substack{m=1 \\ m\ne l}}^{NS}\exp[i{\bf q}\cdot({\bf r}_{CM}^{l}-{\bf r}_{CM}^{m})]><F_{l}({\bf q})F_{m}^{*}({\bf q})>\nonumber\\
\end{eqnarray}

Defining the center-to-center structure factor $S(q)$ 
\begin{equation}
S({\bf q})-1\equiv\frac{1}{N_{S}}<\sum_{l=1}^{N_{S}}\sum_{\substack{m=1 \\ m\ne l}}^{N_{S}}\exp[i{\bf q}\cdot({\bf r}_{_{CM}}^{l}-{\bf r}_{_{CM}}^{m})]>
\end{equation}
and considering the previous definitions for $P({\bf q})$ and $F({\bf q})$ one
can write 
\begin{equation}
I({\bf q})=P({\bf q})+[S({\bf q})-1]F({\bf q})\label{eq:nonclassic}
\end{equation}

In the limit of non-interacting particles, $S({\bf q})=1$ and $I({\bf q})$ provides
a measure of the form factor $P({\bf q})$. The measured form factor is
sufficient to evaluate $S({\bf q})$, if the approximation $F({\bf q})=P({\bf q})$
is valid in the region where $S({\bf q})\ne1$. In this further case 
\begin{equation}
I({\bf q})=S({\bf q})P({\bf q})\label{eq:classic}
\end{equation}

Since we have access to the atomic coordinates in the numerical study,
we can evaluate both $P({\bf q})$ and $F({\bf q})$, both reported in Fig.~\ref{fig:form},  to be used in
conjunction with the experimental $I({\bf q})$ for extracting $S({\bf q})$,
according to Eq.~\ref{eq:nonclassic} and Eq.~\ref{eq:classic}.
The resulting structure factor has to be considered as an experimentally
accessible effective structure factor, since it still depends on the
decoupling approximation between positions and orientation and the
decoupling of the orientation between different pairs. For both routes
(Eq.~\ref{eq:nonclassic} and Eq.~\ref{eq:classic}) we find a consistent
$S({\bf q})$ prediction, even if the
approximation $S({\bf q})=I({\bf q})/P({\bf q})$ is preferred since it does not suffer
from numerical errors associated to the ratio between small numbers,
encountered at large $q$ when using $S({\bf q})=1+[I({\bf q})-P({\bf q})]/F({\bf q})$.

\begin{figure}[t]
\vspace*{-0.35cm}
\centering{}\includegraphics[scale=0.33]{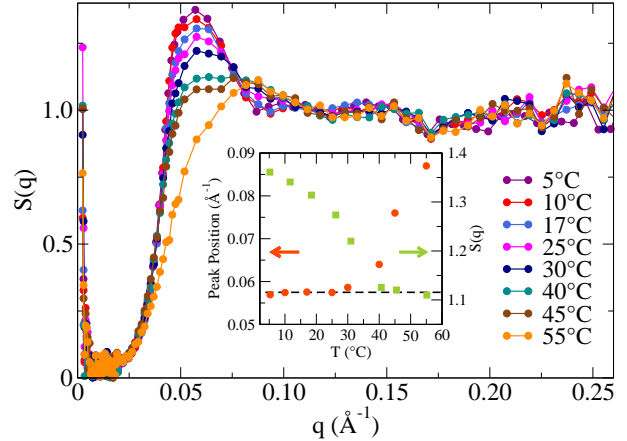}
\caption{\label{fig:sq}Static structure factor $S(q)$ at different temperatures
($55-5$\textdegree C) calculated from the ratio between $I(q)$ and
$P_{sim}(q)$. The shift of the peak position ($\textrm{\AA}^{-1}$)
(dark orange dots) and its intensity (light green squares) as a function of temperature
are displayed in the inset. }
\end{figure}

Fig.~\ref{fig:sq} shows the effective structure factor,  
$$
S(q)=\frac{I(q)}{\beta P(q)}
$$ 
where, once more, $P(q)$ is the normalized form factor and $\beta$ the form factor value at the origin.
The $T$-dependence of the effective structure factor shows the onset
of a peak at a $q^{peak}$ position that shifts to lower $q$ values
(see inset of Fig.~\ref{fig:sq}) as the temperature decreases until
it stabilizes at $q^{peak}=0.0585\textrm{ \AA}^{-1}$ ($2\pi/q^{peak}=107.4\textrm{ \AA}$)
once the gel is formed. $S(q^{peak})$ increases slower and slower
on cooling. This behavior, predicted theoretically and supported
by numerical calculations has been observed in the ageing process
of a clay gel~\cite{ruzicka2011observation} but never observed experimentally
in a controlled and designed thermoreversible system. In the theoretical
studies, saturation results from the formation of a fully bonded network
of tetravalent particles. In this \textquotedbl{}ground zero\textquotedbl{}
structure, all possible bonds are essentially formed and structural
evolution is completed. Dynamics is still possible via the rare breaking
and reforming of the inter-NS bonds on a timescale dictated by the
free-energy cost $\Delta G_{breaking}$. This last quantity has been
estimated by DLS experiments to be about $1.3~\Delta G_{CGATCG}$
where $\Delta G_{CGATCG}$ is the known~\cite{santalucia1998unified}
binding free energy of the sticky CGATCG sequence. At the lowest investigate
$T$, the bond-breaking timescale is larger than 10 seconds. Fig.~\ref{fig:sq}
also shows that the system has a very small compressibility (the limit
of $S(q)$ for $q\rightarrow0$), as expected for a solution of significantly
charged particle  at small ionic strength~\cite{van2008introduction}.

\section{Conclusions}

We have investigated the aggregation process of self-assembled DNA
NSs, generated by mixing equimolar quantities of four properly designed $49$-bases 
DNA oligomers. On cooling, these four strands first
associate to form a four-arms star with sticky ends, followed by aggregation
of these four-functional supramolecules in a thermoreversible gel
structure. The dynamics and phase behavior of this interesting biomaterial
have been previously investigated via light scattering. 

Here, we provided a  structural characterization of the system, 
resulting from  to the  synergy between experiments and computer simulations.

Specifically, we reported the first SANS measurement of the form factor
and compared it with predictions based on oxDNA2~\cite{snodin2015introducing},
a recently developed coarse grained model for investigating DNA nanotechnology.
The predictions of the model are found to be in very good agreement
with the experimental results in the entire wavevector range, allowing
us to estimate precisely the shape of the NS. For the
investigated low salt concentration, the NS is found to be rather planar,
a geometry that minimizes the electrostatic repulsions. Simulations
also provide evidence that the $T$ effect on the shape is negligible.
By contrast, the ionic strength exerts a strong dependence on the shape of the NS.
On increasing salt concentration, the gyration radius significantly
decreases and the four arms fluctuate more freely.

\smallskip{}

SANS measurements in the equilibrium gel region showed the presence of
a peak in the scattered intensity whose amplitude evolves during the
aggregation process and appears to level off at the lowest investigated
$T$s, suggesting that a structurally complete fully bonded tetrahedral
network has been formed. The effective structure factor, evaluated
assuming the validity of the relation $S(q)=I(q)/P(q)$, confirms
the previous analysis, in agreement with theoretical predictions of
simplified colloidal patchy-particle models~\cite{sciortino2011reversible}.
Unfortunately, low ionic-strength simulations of the aggregation process
with an accurate DNA model, even at the coarse grained oxDNA2 level,
are still unfeasible and a direct comparison between simulations and
experiments in the gel phase is still lacking. Hopefully, a new experiment
at significantly large ionic strength (where bulk simulations start
to be feasible~\cite{2014_Accurate_phase_diagram_4valent_DNA-NS})
will allow us to clarify the dependence of the structural properties
and the ionic strength effect on the gel structure. Additional experiments
could also assess the validity of the decoupling between
translational and orientational correlations and substantiate the
interpretation of the effective $S(q)$ as the center-to-center structure
factor.

SANS techniques have often been applied
to the study of gels composed by polymers \cite{2008_Definition_PolymerGels,2014_Supramolecular_polymer_gels,2007_Smart_Polymeric_Gels}
(mostly irreversibly cross-linked~\cite{2009_Shibayama_Matsunaga,1989_BOOK_Lee_NewTrends,1976_Benoit_PolymNetworks,1982_Ullman_Macromolecules,1988_Tsay_Ullman}),
proteins \cite{2012_Peptide_protein_hydrogels,2005_Genetically_engineered_protein_hydrogels,2015_Associative_Protein_Hydrogels}
or nanocomposite clays \cite{2008_PAAMlaponite_Clay_Nanocomposite,2010_PEG_Clay_nanocomposite_hydrogel,2015_Clay_Mineral_Nanocomposite_hydrogels,ruzicka2011observation},
but only rarely interpreted in terms of form and structure factors
even for chemical gels very similar to the present physical one (e.g.
the binary system developed by the group of Shibayama based on two
four-arm poly(ethylene glycol) polymers~\cite{2008_Sakai_Shibayama,2011_Shibayama}).
The DNA gel discussed here, built on the base-pair selectivity,
represents a realisation of an ideal biocompatible physical gel free
of entanglement and defects and with a well defined supramolecular
unit. Such monodisperse constituents of the network nodes, together
with the possibility to compare the neutron data with the accurate
geometry provided by the simulations, is crucial for evaluating an
effective gel structure factor.

\subsection*{Acknowledgments}

FS and JFC acknowledge support from ETN-COLLDENSE (H2020-MCSA-ITN-2014,
Grant No. 642774). FS, TB, FB acknowledge support from MIUR-PRIN. LR acknowledges support
from the Austrian Research Fund (FWF) through the Lise-Meitner Fellowship M 1650-N27,
and from the European Commission through the Marie Sk\l{}odowska-Curie Fellowship 702298-DELTAS.

We also acknowledge ILL, LLB and the HZB for beamtime allocation
and the support on data analysis. This project has received funding from the
European Union's Seventh Framework Programme for research, technological development
and demonstration under the NMI3-II Grant number 283883.

\subsection*{Additional Information}
Competing financial interests: the authors declare no competing financial interests.

\bibliography{Form_Factor_Gels_Bib}

\begin{thebibliography}{77}%
\makeatletter
\providecommand \@ifxundefined [1]{%
 \@ifx{#1\undefined}
}%
\providecommand \@ifnum [1]{%
 \ifnum #1\expandafter \@firstoftwo
 \else \expandafter \@secondoftwo
 \fi
}%
\providecommand \@ifx [1]{%
 \ifx #1\expandafter \@firstoftwo
 \else \expandafter \@secondoftwo
 \fi
}%
\providecommand \natexlab [1]{#1}%
\providecommand \enquote  [1]{``#1''}%
\providecommand \bibnamefont  [1]{#1}%
\providecommand \bibfnamefont [1]{#1}%
\providecommand \citenamefont [1]{#1}%
\providecommand \href@noop [0]{\@secondoftwo}%
\providecommand \href [0]{\begingroup \@sanitize@url \@href}%
\providecommand \@href[1]{\@@startlink{#1}\@@href}%
\providecommand \@@href[1]{\endgroup#1\@@endlink}%
\providecommand \@sanitize@url [0]{\catcode `\\12\catcode `\$12\catcode
  `\&12\catcode `\#12\catcode `\^12\catcode `\_12\catcode `\%12\relax}%
\providecommand \@@startlink[1]{}%
\providecommand \@@endlink[0]{}%
\providecommand \url  [0]{\begingroup\@sanitize@url \@url }%
\providecommand \@url [1]{\endgroup\@href {#1}{\urlprefix }}%
\providecommand \urlprefix  [0]{URL }%
\providecommand \Eprint [0]{\href }%
\providecommand \doibase [0]{http://dx.doi.org/}%
\providecommand \selectlanguage [0]{\@gobble}%
\providecommand \bibinfo  [0]{\@secondoftwo}%
\providecommand \bibfield  [0]{\@secondoftwo}%
\providecommand \translation [1]{[#1]}%
\providecommand \BibitemOpen [0]{}%
\providecommand \bibitemStop [0]{}%
\providecommand \bibitemNoStop [0]{.\EOS\space}%
\providecommand \EOS [0]{\spacefactor3000\relax}%
\providecommand \BibitemShut  [1]{\csname bibitem#1\endcsname}%
\let\auto@bib@innerbib\@empty
\bibitem [{\citenamefont {{M. Yoshida, and J.
  Lahann}}(2008)}]{2008__APP_SmartMaterials}%
  \BibitemOpen
  \bibfield  {author} {\bibinfo {author} {\bibnamefont {{M. Yoshida, and J.
  Lahann}}},\ }\href@noop {} {\bibfield  {journal} {\bibinfo  {journal} {{ACS
  Nano}}\ }\textbf {\bibinfo {volume} {26}},\ \bibinfo {pages} {1101--1107}
  (\bibinfo {year} {2008})}\BibitemShut {NoStop}%
\bibitem [{\citenamefont {{A. Condon}}(2006)}]{2006_APP_MolecularNanotech}%
  \BibitemOpen
  \bibfield  {author} {\bibinfo {author} {\bibnamefont {{A. Condon}}},\
  }\href@noop {} {\bibfield  {journal} {\bibinfo  {journal} {{Nat. Rev.
  Genet.}}\ }\textbf {\bibinfo {volume} {7}},\ \bibinfo {pages} {565--575}
  (\bibinfo {year} {2006})}\BibitemShut {NoStop}%
\bibitem [{\citenamefont {{M. Patil, D. S. Mehta, and S.
  Guvva}}(2008)}]{2008_Impact_Nanotechn_medicine_dentistry}%
  \BibitemOpen
  \bibfield  {author} {\bibinfo {author} {\bibnamefont {{M. Patil, D. S. Mehta,
  and S. Guvva}}},\ }\href@noop {} {\bibfield  {journal} {\bibinfo  {journal}
  {{J. Indian Soc. Periodontol.}}\ }\textbf {\bibinfo {volume} {12}} (\bibinfo
  {year} {2008})}\BibitemShut {NoStop}%
\bibitem [{\citenamefont {{P. C. Nicolson, and J.
  Vogh}}(2001)}]{2001_APP_ContactLenses}%
  \BibitemOpen
  \bibfield  {author} {\bibinfo {author} {\bibnamefont {{P. C. Nicolson, and J.
  Vogh}}},\ }\href@noop {} {\bibfield  {journal} {\bibinfo  {journal}
  {{Biomaterials}}\ }\textbf {\bibinfo {volume} {22}},\ \bibinfo {pages}
  {3273--3283} (\bibinfo {year} {2001})}\BibitemShut {NoStop}%
\bibitem [{\citenamefont {{N. A. Peppas, J. Z. Hilt, A. Khademhosseini, and R.
  Langer}}(2006)}]{2006_APP_Medicine_Pepas}%
  \BibitemOpen
  \bibfield  {author} {\bibinfo {author} {\bibnamefont {{N. A. Peppas, J. Z.
  Hilt, A. Khademhosseini, and R. Langer}}},\ }\href@noop {} {\bibfield
  {journal} {\bibinfo  {journal} {{Adv. Mater.}}\ }\textbf {\bibinfo {volume}
  {18}},\ \bibinfo {pages} {1345--1360} (\bibinfo {year} {2006})}\BibitemShut
  {NoStop}%
\bibitem [{\citenamefont {{J. Kope{\v c}ek}}(2003)}]{2003_APP_Kopecek_DrugDel}%
  \BibitemOpen
  \bibfield  {author} {\bibinfo {author} {\bibnamefont {{J. Kope{\v c}ek}}},\
  }\href@noop {} {\bibfield  {journal} {\bibinfo  {journal} {{Eur. J. Pharm.
  Sci.}}\ }\textbf {\bibinfo {volume} {20}},\ \bibinfo {pages} {1--16}
  (\bibinfo {year} {2003})}\BibitemShut {NoStop}%
\bibitem [{\citenamefont {{D. Costa, A. J. M: Valente, M. Gra{\c c}a Miguel,
  and J. Queiroz}}(2014)}]{2014_APP_CancerTherapy_Portugal}%
  \BibitemOpen
  \bibfield  {author} {\bibinfo {author} {\bibnamefont {{D. Costa, A. J. M:
  Valente, M. Gra{\c c}a Miguel, and J. Queiroz}}},\ }\href@noop {} {\bibfield
  {journal} {\bibinfo  {journal} {{Colloids Surf., A}}\ }\textbf {\bibinfo
  {volume} {442}},\ \bibinfo {pages} {181--190} (\bibinfo {year}
  {2014})}\BibitemShut {NoStop}%
\bibitem [{\citenamefont {{D. Tada, T. Tanabe, A. Tachibana, and K.
  Yamauchi}}(2005)}]{2005_APP_DrugDel_Tachibana}%
  \BibitemOpen
  \bibfield  {author} {\bibinfo {author} {\bibnamefont {{D. Tada, T. Tanabe, A.
  Tachibana, and K. Yamauchi}}},\ }\href@noop {} {\bibfield  {journal}
  {\bibinfo  {journal} {{J. Biosci. Bioeng.}}\ }\textbf {\bibinfo {volume}
  {100}},\ \bibinfo {pages} {551--555} (\bibinfo {year} {2005})}\BibitemShut
  {NoStop}%
\bibitem [{\citenamefont {{A. J.
  Steckl}}(2007)}]{2007_APP_Photonics_NewMaterial}%
  \BibitemOpen
  \bibfield  {author} {\bibinfo {author} {\bibnamefont {{A. J. Steckl}}},\
  }\href@noop {} {\bibfield  {journal} {\bibinfo  {journal} {{Nature Photon.}}\
  }\textbf {\bibinfo {volume} {1}},\ \bibinfo {pages} {3--5} (\bibinfo {year}
  {2007})}\BibitemShut {NoStop}%
\bibitem [{\citenamefont {{A. J. Steckl, H. Spaeth, H. You, E. Gomez, and J.
  Grote}}(2011)}]{2011_APP_Photonics_DNA}%
  \BibitemOpen
  \bibfield  {author} {\bibinfo {author} {\bibnamefont {{A. J. Steckl, H.
  Spaeth, H. You, E. Gomez, and J. Grote}}},\ }\href@noop {} {\bibfield
  {journal} {\bibinfo  {journal} {{Opt. Photonics News}}\ }\textbf {\bibinfo
  {volume} {22}},\ \bibinfo {pages} {34--39} (\bibinfo {year}
  {2011})}\BibitemShut {NoStop}%
\bibitem [{\citenamefont {{D. Boneh, C. Dunworth, R. J. Lipton, and J. I.
  Sgall}}(1996)}]{1996_APP_Computing_Turing1}%
  \BibitemOpen
  \bibfield  {author} {\bibinfo {author} {\bibnamefont {{D. Boneh, C. Dunworth,
  R. J. Lipton, and J. I. Sgall}}},\ }\href@noop {} {\bibfield  {journal}
  {\bibinfo  {journal} {{Discrete Appl. Math.}}\ }\textbf {\bibinfo {volume}
  {71}} (\bibinfo {year} {1996})}\BibitemShut {NoStop}%
\bibitem [{\citenamefont {{S. M. Douglas, I. Bachelet, and G. M.
  Church}}(2012)}]{2012_APP_Computing_nanorobots}%
  \BibitemOpen
  \bibfield  {author} {\bibinfo {author} {\bibnamefont {{S. M. Douglas, I.
  Bachelet, and G. M. Church}}},\ }\href@noop {} {\bibfield  {journal}
  {\bibinfo  {journal} {{Science}}\ }\textbf {\bibinfo {volume} {335}},\
  \bibinfo {pages} {831--834} (\bibinfo {year} {2012})}\BibitemShut {NoStop}%
\bibitem [{\citenamefont {{N. C.
  Seeman}}(2003)}]{2003_Seeman_DNA_Material_World}%
  \BibitemOpen
  \bibfield  {author} {\bibinfo {author} {\bibnamefont {{N. C. Seeman}}},\
  }\href@noop {} {\bibfield  {journal} {\bibinfo  {journal} {{Nature}}\
  }\textbf {\bibinfo {volume} {421}},\ \bibinfo {pages} {427--431} (\bibinfo
  {year} {2003})}\BibitemShut {NoStop}%
\bibitem [{\citenamefont {{Y. W. Kwon, C. H. Lee, D. H. Choi, and J.Il
  Jin}}(2009)}]{2009_APP_MaterialsScienceDNA}%
  \BibitemOpen
  \bibfield  {author} {\bibinfo {author} {\bibnamefont {{Y. W. Kwon, C. H. Lee,
  D. H. Choi, and J.Il Jin}}},\ }\href@noop {} {\bibfield  {journal} {\bibinfo
  {journal} {{J. Mater. Chem.}}\ }\textbf {\bibinfo {volume} {19}},\ \bibinfo
  {pages} {1353--1380} (\bibinfo {year} {2009})}\BibitemShut {NoStop}%
\bibitem [{\citenamefont {Zadegan}\ and\ \citenamefont
  {Norton}(2012)}]{zadegan2012structural}%
  \BibitemOpen
  \bibfield  {author} {\bibinfo {author} {\bibfnamefont {R.~M.}\ \bibnamefont
  {Zadegan}}\ and\ \bibinfo {author} {\bibfnamefont {M.~L.}\ \bibnamefont
  {Norton}},\ }\href@noop {} {\bibfield  {journal} {\bibinfo  {journal} {Int.
  J. Mol. Sci.}\ }\textbf {\bibinfo {volume} {13}},\ \bibinfo {pages}
  {7149--7162} (\bibinfo {year} {2012})}\BibitemShut {NoStop}%
\bibitem [{\citenamefont {Biffi}\ \emph {et~al.}(2013)\citenamefont {Biffi},
  \citenamefont {Cerbino}, \citenamefont {Bomboi}, \citenamefont {Paraboschi},
  \citenamefont {Asselta}, \citenamefont {Sciortino},\ and\ \citenamefont
  {Bellini}}]{biffi2013phase}%
  \BibitemOpen
  \bibfield  {author} {\bibinfo {author} {\bibfnamefont {S.}~\bibnamefont
  {Biffi}}, \bibinfo {author} {\bibfnamefont {R.}~\bibnamefont {Cerbino}},
  \bibinfo {author} {\bibfnamefont {F.}~\bibnamefont {Bomboi}}, \bibinfo
  {author} {\bibfnamefont {E.~M.}\ \bibnamefont {Paraboschi}}, \bibinfo
  {author} {\bibfnamefont {R.}~\bibnamefont {Asselta}}, \bibinfo {author}
  {\bibfnamefont {F.}~\bibnamefont {Sciortino}}, \ and\ \bibinfo {author}
  {\bibfnamefont {T.}~\bibnamefont {Bellini}},\ }\href@noop {} {\bibfield
  {journal} {\bibinfo  {journal} {Proc. Natl. Acad. Sci. USA}\ }\textbf
  {\bibinfo {volume} {110}},\ \bibinfo {pages} {15633--15637} (\bibinfo {year}
  {2013})}\BibitemShut {NoStop}%
\bibitem [{\citenamefont {{S. Biffi, R. Cerbino, G. Nava, F. Bomboi, F.
  Sciortino, and T.
  Bellini}}(2015)}]{2015_Equilibrium_gels_DNA_nanostars_Bomboi}%
  \BibitemOpen
  \bibfield  {author} {\bibinfo {author} {\bibnamefont {{S. Biffi, R. Cerbino,
  G. Nava, F. Bomboi, F. Sciortino, and T. Bellini}}},\ }\href@noop {}
  {\bibfield  {journal} {\bibinfo  {journal} {{Soft Matter}}\ }\textbf
  {\bibinfo {volume} {11}} (\bibinfo {year} {2015})}\BibitemShut {NoStop}%
\bibitem [{\citenamefont {Bianchi}\ \emph {et~al.}(2006)\citenamefont
  {Bianchi}, \citenamefont {Largo}, \citenamefont {Tartaglia}, \citenamefont
  {Zaccarelli},\ and\ \citenamefont {Sciortino}}]{bianchi2006phase}%
  \BibitemOpen
  \bibfield  {author} {\bibinfo {author} {\bibfnamefont {E.}~\bibnamefont
  {Bianchi}}, \bibinfo {author} {\bibfnamefont {J.}~\bibnamefont {Largo}},
  \bibinfo {author} {\bibfnamefont {P.}~\bibnamefont {Tartaglia}}, \bibinfo
  {author} {\bibfnamefont {E.}~\bibnamefont {Zaccarelli}}, \ and\ \bibinfo
  {author} {\bibfnamefont {F.}~\bibnamefont {Sciortino}},\ }\href@noop {}
  {\bibfield  {journal} {\bibinfo  {journal} {Phys. Rev. Lett.}\ }\textbf
  {\bibinfo {volume} {97}},\ \bibinfo {pages} {168301} (\bibinfo {year}
  {2006})}\BibitemShut {NoStop}%
\bibitem [{\citenamefont {{P. J. Lu, E. Zaccarelli, F. Ciulla, A. B. Schofield,
  F. Sciortino, and D. A. Weitz}}(2008)}]{2008_Zaccarelli_Sciortino}%
  \BibitemOpen
  \bibfield  {author} {\bibinfo {author} {\bibnamefont {{P. J. Lu, E.
  Zaccarelli, F. Ciulla, A. B. Schofield, F. Sciortino, and D. A. Weitz}}},\
  }\href@noop {} {\bibfield  {journal} {\bibinfo  {journal} {{Nature}}\
  }\textbf {\bibinfo {volume} {453}} (\bibinfo {year} {2008})}\BibitemShut
  {NoStop}%
\bibitem [{\citenamefont {{T. Sakai, T. Matsunaga, Y. Yamamoto, C. Ito, R.
  Yoshida, S. Suzuki, N. Sasaki, and M.
  Shibayama}}(2008)}]{2008_Sakai_Shibayama}%
  \BibitemOpen
  \bibfield  {author} {\bibinfo {author} {\bibnamefont {{T. Sakai, T.
  Matsunaga, Y. Yamamoto, C. Ito, R. Yoshida, S. Suzuki, N. Sasaki, and M.
  Shibayama}}},\ }\href@noop {} {\bibfield  {journal} {\bibinfo  {journal}
  {{Macromolecules}}\ }\textbf {\bibinfo {volume} {41}},\ \bibinfo {pages}
  {5379--5384} (\bibinfo {year} {2008})}\BibitemShut {NoStop}%
\bibitem [{\citenamefont {Bomboi}\ \emph {et~al.}(2015)\citenamefont {Bomboi},
  \citenamefont {Biffi}, \citenamefont {Cerbino}, \citenamefont {Bellini},
  \citenamefont {Bordi},\ and\ \citenamefont
  {Sciortino}}]{bomboi2015equilibrium}%
  \BibitemOpen
  \bibfield  {author} {\bibinfo {author} {\bibfnamefont {F.}~\bibnamefont
  {Bomboi}}, \bibinfo {author} {\bibfnamefont {S.}~\bibnamefont {Biffi}},
  \bibinfo {author} {\bibfnamefont {R.}~\bibnamefont {Cerbino}}, \bibinfo
  {author} {\bibfnamefont {T.}~\bibnamefont {Bellini}}, \bibinfo {author}
  {\bibfnamefont {F.}~\bibnamefont {Bordi}}, \ and\ \bibinfo {author}
  {\bibfnamefont {F.}~\bibnamefont {Sciortino}},\ }\href@noop {} {\bibfield
  {journal} {\bibinfo  {journal} {Eur. Phys. J. E Soft Matter}\ }\textbf
  {\bibinfo {volume} {38}},\ \bibinfo {pages} {1--8} (\bibinfo {year}
  {2015})}\BibitemShut {NoStop}%
\bibitem [{\citenamefont {Doye}\ \emph {et~al.}(2013)\citenamefont {Doye},
  \citenamefont {Ouldridge}, \citenamefont {Louis}, \citenamefont {Romano},
  \citenamefont {{\v{S}}ulc}, \citenamefont {Matek}, \citenamefont {Snodin},
  \citenamefont {Rovigatti}, \citenamefont {Schreck}, \citenamefont {Harrison}
  \emph {et~al.}}]{doye2013coarse}%
  \BibitemOpen
  \bibfield  {author} {\bibinfo {author} {\bibfnamefont {J.~P.}\ \bibnamefont
  {Doye}}, \bibinfo {author} {\bibfnamefont {T.~E.}\ \bibnamefont {Ouldridge}},
  \bibinfo {author} {\bibfnamefont {A.~A.}\ \bibnamefont {Louis}}, \bibinfo
  {author} {\bibfnamefont {F.}~\bibnamefont {Romano}}, \bibinfo {author}
  {\bibfnamefont {P.}~\bibnamefont {{\v{S}}ulc}}, \bibinfo {author}
  {\bibfnamefont {C.}~\bibnamefont {Matek}}, \bibinfo {author} {\bibfnamefont
  {B.~E.}\ \bibnamefont {Snodin}}, \bibinfo {author} {\bibfnamefont
  {L.}~\bibnamefont {Rovigatti}}, \bibinfo {author} {\bibfnamefont {J.~S.}\
  \bibnamefont {Schreck}}, \bibinfo {author} {\bibfnamefont {R.~M.}\
  \bibnamefont {Harrison}},  \emph {et~al.},\ }\href@noop {} {\bibfield
  {journal} {\bibinfo  {journal} {Phys. Chem. Chem. Phys.}\ }\textbf {\bibinfo
  {volume} {15}},\ \bibinfo {pages} {20395--20414} (\bibinfo {year}
  {2013})}\BibitemShut {NoStop}%
\bibitem [{\citenamefont {{P. Desjardins, and D.
  Conklin}}(2010)}]{2010_Nanodrop}%
  \BibitemOpen
  \bibfield  {author} {\bibinfo {author} {\bibnamefont {{P. Desjardins, and D.
  Conklin}}},\ }\href@noop {} {\bibfield  {journal} {\bibinfo  {journal} {{J.
  Vis. Exp.}}\ }\textbf {\bibinfo {volume} {45}} (\bibinfo {year}
  {2010})}\BibitemShut {NoStop}%
\bibitem [{\citenamefont {{C. D. Dewhurst}}(2008)}]{2008_D33_ILL}%
  \BibitemOpen
  \bibfield  {author} {\bibinfo {author} {\bibnamefont {{C. D. Dewhurst}}},\
  }\href@noop {} {\bibfield  {journal} {\bibinfo  {journal} {{Meas. Sci.
  Technol.}}\ }\textbf {\bibinfo {volume} {19}} (\bibinfo {year}
  {2008})}\BibitemShut {NoStop}%
\bibitem [{htt()}]{http_www_ill_fr_d22}%
  \BibitemOpen
  \href@noop {} {}\bibinfo {howpublished}
  {\url{http://www.ill.fr/D22/}}\BibitemShut {NoStop}%
\bibitem [{\citenamefont {{C. Dewhurst}}(2003)}]{2003__GRASP_Manual}%
  \BibitemOpen
  \bibfield  {author} {\bibinfo {author} {\bibnamefont {{C. Dewhurst}}},\
  }\href@noop {} {}\bibinfo {organization} {{ILL}} (\bibinfo {year}
  {2003})\BibitemShut {NoStop}%
\bibitem [{\citenamefont {{U. Keiderling, and A.
  Wiedenmann}}(1995)}]{V4_BerII_Berlin}%
  \BibitemOpen
  \bibfield  {author} {\bibinfo {author} {\bibnamefont {{U. Keiderling, and A.
  Wiedenmann}}},\ }\href@noop {} {\bibfield  {journal} {\bibinfo  {journal}
  {{Physica B}}\ }\textbf {\bibinfo {volume} {{213-214}}},\ \bibinfo {pages}
  {895--897} (\bibinfo {year} {1995})}\BibitemShut {NoStop}%
\bibitem [{\citenamefont {{U. Keiderling}}(2002)}]{2002_BerSANS_PC}%
  \BibitemOpen
  \bibfield  {author} {\bibinfo {author} {\bibnamefont {{U. Keiderling}}},\
  }\href@noop {} {\bibfield  {journal} {\bibinfo  {journal} {{Appl. Phys. A}}\
  }\textbf {\bibinfo {volume} {74}} (\bibinfo {year} {2002})}\BibitemShut
  {NoStop}%
\bibitem [{\citenamefont {Snodin}\ \emph {et~al.}(2015)\citenamefont {Snodin},
  \citenamefont {Randisi}, \citenamefont {Mosayebi}, \citenamefont
  {{\v{S}}ulc}, \citenamefont {Schreck}, \citenamefont {Romano}, \citenamefont
  {Ouldridge}, \citenamefont {Tsukanov}, \citenamefont {Nir}, \citenamefont
  {Louis} \emph {et~al.}}]{snodin2015introducing}%
  \BibitemOpen
  \bibfield  {author} {\bibinfo {author} {\bibfnamefont {B.~E.}\ \bibnamefont
  {Snodin}}, \bibinfo {author} {\bibfnamefont {F.}~\bibnamefont {Randisi}},
  \bibinfo {author} {\bibfnamefont {M.}~\bibnamefont {Mosayebi}}, \bibinfo
  {author} {\bibfnamefont {P.}~\bibnamefont {{\v{S}}ulc}}, \bibinfo {author}
  {\bibfnamefont {J.~S.}\ \bibnamefont {Schreck}}, \bibinfo {author}
  {\bibfnamefont {F.}~\bibnamefont {Romano}}, \bibinfo {author} {\bibfnamefont
  {T.~E.}\ \bibnamefont {Ouldridge}}, \bibinfo {author} {\bibfnamefont
  {R.}~\bibnamefont {Tsukanov}}, \bibinfo {author} {\bibfnamefont
  {E.}~\bibnamefont {Nir}}, \bibinfo {author} {\bibfnamefont {A.~A.}\
  \bibnamefont {Louis}},  \emph {et~al.},\ }\href@noop {} {\bibfield  {journal}
  {\bibinfo  {journal} {J. Chem. Phys.}\ }\textbf {\bibinfo {volume} {142}},\
  \bibinfo {pages} {234901} (\bibinfo {year} {2015})}\BibitemShut {NoStop}%
\bibitem [{\citenamefont {Ouldridge}, \citenamefont {Louis},\ and\
  \citenamefont {Doye}(2011)}]{ouldridge2011structural}%
  \BibitemOpen
  \bibfield  {author} {\bibinfo {author} {\bibfnamefont {T.~E.}\ \bibnamefont
  {Ouldridge}}, \bibinfo {author} {\bibfnamefont {A.~A.}\ \bibnamefont
  {Louis}}, \ and\ \bibinfo {author} {\bibfnamefont {J.~P.}\ \bibnamefont
  {Doye}},\ }\href@noop {} {\bibfield  {journal} {\bibinfo  {journal} {J. Chem.
  Phys.}\ }\textbf {\bibinfo {volume} {134}},\ \bibinfo {pages} {085101}
  (\bibinfo {year} {2011})}\BibitemShut {NoStop}%
\bibitem [{\citenamefont {{\v{S}}ulc}\ \emph {et~al.}(2012)\citenamefont
  {{\v{S}}ulc}, \citenamefont {Romano}, \citenamefont {Ouldridge},
  \citenamefont {Rovigatti}, \citenamefont {Doye},\ and\ \citenamefont
  {Louis}}]{vsulc2012sequence}%
  \BibitemOpen
  \bibfield  {author} {\bibinfo {author} {\bibfnamefont {P.}~\bibnamefont
  {{\v{S}}ulc}}, \bibinfo {author} {\bibfnamefont {F.}~\bibnamefont {Romano}},
  \bibinfo {author} {\bibfnamefont {T.~E.}\ \bibnamefont {Ouldridge}}, \bibinfo
  {author} {\bibfnamefont {L.}~\bibnamefont {Rovigatti}}, \bibinfo {author}
  {\bibfnamefont {J.~P.}\ \bibnamefont {Doye}}, \ and\ \bibinfo {author}
  {\bibfnamefont {A.~A.}\ \bibnamefont {Louis}},\ }\href@noop {} {\bibfield
  {journal} {\bibinfo  {journal} {J. Chem. Phys.}\ }\textbf {\bibinfo {volume}
  {137}},\ \bibinfo {pages} {135101} (\bibinfo {year} {2012})}\BibitemShut
  {NoStop}%
\bibitem [{nis()}]{nist}%
  \BibitemOpen
  \href@noop {} {}\bibinfo {howpublished}
  {\url{https://www.ncnr.nist.gov/resources/n-lengths/}}\BibitemShut {NoStop}%
\bibitem [{\citenamefont {Chen}\ and\ \citenamefont
  {Tartaglia}(2015)}]{chen2015scattering}%
  \BibitemOpen
  \bibfield  {author} {\bibinfo {author} {\bibfnamefont {S.-H.}\ \bibnamefont
  {Chen}}\ and\ \bibinfo {author} {\bibfnamefont {P.}~\bibnamefont
  {Tartaglia}},\ }\href@noop {} {\emph {\bibinfo {title} {Scattering Methods in
  Complex Fluids}}}\ (\bibinfo  {publisher} {Cambridge University Press},\
  \bibinfo {year} {2015})\BibitemShut {NoStop}%
\bibitem [{\citenamefont {Kotlarchyk}\ and\ \citenamefont
  {Chen}(1983)}]{kotlarchyk1983analysis}%
  \BibitemOpen
  \bibfield  {author} {\bibinfo {author} {\bibfnamefont {M.}~\bibnamefont
  {Kotlarchyk}}\ and\ \bibinfo {author} {\bibfnamefont {S.-H.}\ \bibnamefont
  {Chen}},\ }\href@noop {} {\bibfield  {journal} {\bibinfo  {journal} {J. Chem.
  Phys.}\ }\textbf {\bibinfo {volume} {79}},\ \bibinfo {pages} {2461--2469}
  (\bibinfo {year} {1983})}\BibitemShut {NoStop}%
\bibitem [{\citenamefont {Jones}(2002)}]{jones2002soft}%
  \BibitemOpen
  \bibfield  {author} {\bibinfo {author} {\bibfnamefont {R.~A.}\ \bibnamefont
  {Jones}},\ }\href@noop {} {\emph {\bibinfo {title} {Soft condensed
  matter}}},\ Vol.~\bibinfo {volume} {6}\ (\bibinfo  {publisher} {Oxford
  University Press},\ \bibinfo {year} {2002})\BibitemShut {NoStop}%
\bibitem [{\citenamefont {Kassapidou}\ \emph {et~al.}(1997)\citenamefont
  {Kassapidou}, \citenamefont {Jesse}, \citenamefont {Kuil}, \citenamefont
  {Lapp}, \citenamefont {Egelhaaf},\ and\ \citenamefont {Van~der
  Maarel}}]{kassapidou1997structure}%
  \BibitemOpen
  \bibfield  {author} {\bibinfo {author} {\bibfnamefont {K.}~\bibnamefont
  {Kassapidou}}, \bibinfo {author} {\bibfnamefont {W.}~\bibnamefont {Jesse}},
  \bibinfo {author} {\bibfnamefont {M.}~\bibnamefont {Kuil}}, \bibinfo {author}
  {\bibfnamefont {A.}~\bibnamefont {Lapp}}, \bibinfo {author} {\bibfnamefont
  {S.}~\bibnamefont {Egelhaaf}}, \ and\ \bibinfo {author} {\bibfnamefont
  {J.}~\bibnamefont {Van~der Maarel}},\ }\href@noop {} {\bibfield  {journal}
  {\bibinfo  {journal} {Macromolecules}\ }\textbf {\bibinfo {volume} {30}},\
  \bibinfo {pages} {2671--2684} (\bibinfo {year} {1997})}\BibitemShut {NoStop}%
\bibitem [{\citenamefont {{A. Guinier, and G.
  Fournet}}()}]{1955_Guinier_Cylinder_Pq}%
  \BibitemOpen
  \bibfield  {author} {\bibinfo {author} {\bibnamefont {{A. Guinier, and G.
  Fournet}}},\ }\href@noop {} {\emph {\bibinfo {title} {Small angle scattering
  of X-rays}}}\BibitemShut {NoStop}%
\bibitem [{\citenamefont {{J. R. C. van der Maarel, and K.
  Kassapidou}}(1998)}]{1998_Short_DNA_fragments}%
  \BibitemOpen
  \bibfield  {author} {\bibinfo {author} {\bibnamefont {{J. R. C. van der
  Maarel, and K. Kassapidou}}},\ }\href@noop {} {\bibfield  {journal} {\bibinfo
   {journal} {{Macromolecules}}\ }\textbf {\bibinfo {volume} {31}},\ \bibinfo
  {pages} {5734--5739} (\bibinfo {year} {1998})}\BibitemShut {NoStop}%
\bibitem [{\citenamefont {{J. R. C. van der
  Maarel}}(1999)}]{1999_Maarel_small_Rc}%
  \BibitemOpen
  \bibfield  {author} {\bibinfo {author} {\bibnamefont {{J. R. C. van der
  Maarel}}},\ }\href@noop {} {\bibfield  {journal} {\bibinfo  {journal}
  {{Biophys. J.}}\ }\textbf {\bibinfo {volume} {76}},\ \bibinfo {pages}
  {2673--2678} (\bibinfo {year} {1999})}\BibitemShut {NoStop}%
\bibitem [{\citenamefont {{H. Lederer, R. P. May, J. K. Kjems, and H.
  Heumann}}(1986)}]{1986_Short_DNA_fragment_SANS}%
  \BibitemOpen
  \bibfield  {author} {\bibinfo {author} {\bibnamefont {{H. Lederer, R. P. May,
  J. K. Kjems, and H. Heumann}}},\ }\href@noop {} {\bibfield  {journal}
  {\bibinfo  {journal} {{Eur. J. Biochem.}}\ }\textbf {\bibinfo {volume}
  {161}},\ \bibinfo {pages} {191--196} (\bibinfo {year} {1986})}\BibitemShut
  {NoStop}%
\bibitem [{\citenamefont {{J. D. Watson, and F. H. C.
  Crick}}(1953)}]{1953_WatsonCrick_DNA_structure}%
  \BibitemOpen
  \bibfield  {author} {\bibinfo {author} {\bibnamefont {{J. D. Watson, and F.
  H. C. Crick}}},\ }\href@noop {} {\bibfield  {journal} {\bibinfo  {journal}
  {{Nature}}\ }\textbf {\bibinfo {volume} {171}},\ \bibinfo {pages} {737--738}
  (\bibinfo {year} {1953})}\BibitemShut {NoStop}%
\bibitem [{\citenamefont {{R. Borsali, H. Nguyen, and R.
  Pecora}}(1998)}]{1998_SANS_DNA_Polyelectrolyte}%
  \BibitemOpen
  \bibfield  {author} {\bibinfo {author} {\bibnamefont {{R. Borsali, H. Nguyen,
  and R. Pecora}}},\ }\href@noop {} {\bibfield  {journal} {\bibinfo  {journal}
  {{Macromolecules}}\ }\textbf {\bibinfo {volume} {31}},\ \bibinfo {pages}
  {1548--1555} (\bibinfo {year} {1998})}\BibitemShut {NoStop}%
\bibitem [{\citenamefont {{V. Luzzati, F. Masson, A. Mathis, and P.
  Saludjian}}(1976)}]{1976_Luzzati_10A_DNA}%
  \BibitemOpen
  \bibfield  {author} {\bibinfo {author} {\bibnamefont {{V. Luzzati, F. Masson,
  A. Mathis, and P. Saludjian}}},\ }\href@noop {} {\bibfield  {journal}
  {\bibinfo  {journal} {{Biopolymers}}\ }\textbf {\bibinfo {volume} {4}}
  (\bibinfo {year} {1976})}\BibitemShut {NoStop}%
\bibitem [{\citenamefont {{J. Garcia de la Torre, and A.
  Horta}}(1976)}]{1976_GarciaTorre_10A_DNA}%
  \BibitemOpen
  \bibfield  {author} {\bibinfo {author} {\bibnamefont {{J. Garcia de la Torre,
  and A. Horta}}},\ }\href@noop {} {\bibfield  {journal} {\bibinfo  {journal}
  {{J. Phys. Chem.}}\ }\textbf {\bibinfo {volume} {80}} (\bibinfo {year}
  {1976})}\BibitemShut {NoStop}%
\bibitem [{\citenamefont {{M. H. J. Koch, Z. Sayers, P. Sicre, and D.
  Svergun}}(1995)}]{1995_Koch_10A_DNA}%
  \BibitemOpen
  \bibfield  {author} {\bibinfo {author} {\bibnamefont {{M. H. J. Koch, Z.
  Sayers, P. Sicre, and D. Svergun}}},\ }\href@noop {} {\bibfield  {journal}
  {\bibinfo  {journal} {{Macromolecules}}\ }\textbf {\bibinfo {volume} {28}}
  (\bibinfo {year} {1995})}\BibitemShut {NoStop}%
\bibitem [{\citenamefont {Hansen}\ and\ \citenamefont
  {McDonald}(1990)}]{hansen1990theory}%
  \BibitemOpen
  \bibfield  {author} {\bibinfo {author} {\bibfnamefont {J.-P.}\ \bibnamefont
  {Hansen}}\ and\ \bibinfo {author} {\bibfnamefont {I.~R.}\ \bibnamefont
  {McDonald}},\ }\href@noop {} {\emph {\bibinfo {title} {Theory of simple
  liquids}}}\ (\bibinfo  {publisher} {Elsevier},\ \bibinfo {year}
  {1990})\BibitemShut {NoStop}%
\bibitem [{\citenamefont {{M. Milas, M. Rinaudo, R. Duplessix, R. Borsali, and
  P. Lindner}}(1995)}]{1995_Xanthan_upturn}%
  \BibitemOpen
  \bibfield  {author} {\bibinfo {author} {\bibnamefont {{M. Milas, M. Rinaudo,
  R. Duplessix, R. Borsali, and P. Lindner}}},\ }\href@noop {} {\bibfield
  {journal} {\bibinfo  {journal} {{Macromolecules}}\ }\textbf {\bibinfo
  {volume} {28}},\ \bibinfo {pages} {3119--3124} (\bibinfo {year}
  {1995})}\BibitemShut {NoStop}%
\bibitem [{\citenamefont {{N. Arfin, V. K. Aswal, J. Kohlbrecher, and H. B.
  Bohidar}}(2015)}]{2015_Polymer_upturn_India}%
  \BibitemOpen
  \bibfield  {author} {\bibinfo {author} {\bibnamefont {{N. Arfin, V. K. Aswal,
  J. Kohlbrecher, and H. B. Bohidar}}},\ }\href@noop {} {\bibfield  {journal}
  {\bibinfo  {journal} {{Polymer}}\ }\textbf {\bibinfo {volume} {65}},\
  \bibinfo {pages} {175--182} (\bibinfo {year} {2015})}\BibitemShut {NoStop}%
\bibitem [{\citenamefont {{M. Sedlak}}(1996)}]{1996_Upturn_contr_sedlak}%
  \BibitemOpen
  \bibfield  {author} {\bibinfo {author} {\bibnamefont {{M. Sedlak}}},\
  }\href@noop {} {\bibfield  {journal} {\bibinfo  {journal} {{J. Chem. Phys.}}\
  }\textbf {\bibinfo {volume} {105}},\ \bibinfo {pages} {10123--10133}
  (\bibinfo {year} {1996})}\BibitemShut {NoStop}%
\bibitem [{\citenamefont {{M.
  Shibayama}}(1998)}]{1998_Shibayama_upturn_inhomog}%
  \BibitemOpen
  \bibfield  {author} {\bibinfo {author} {\bibnamefont {{M. Shibayama}}},\
  }\href@noop {} {\bibfield  {journal} {\bibinfo  {journal} {{Macrom. Chem.
  Phys.}}\ }\textbf {\bibinfo {volume} {199}},\ \bibinfo {pages} {1--30}
  (\bibinfo {year} {1998})}\BibitemShut {NoStop}%
\bibitem [{\citenamefont {{H. Matsuoka, D. Schwahn, and N.
  Ise}}(1991)}]{1991_Matsuoka_upturn_discussion}%
  \BibitemOpen
  \bibfield  {author} {\bibinfo {author} {\bibnamefont {{H. Matsuoka, D.
  Schwahn, and N. Ise}}},\ }\href@noop {} {\bibfield  {journal} {\bibinfo
  {journal} {{Macromolecules}}\ }\textbf {\bibinfo {volume} {24}} (\bibinfo
  {year} {1991})}\BibitemShut {NoStop}%
\bibitem [{\citenamefont {{Y. Zhang, J. F. Douglas, B. D. Ermi, and E. J.
  Amis}}(2001)}]{2001_upturn_contr_zhang}%
  \BibitemOpen
  \bibfield  {author} {\bibinfo {author} {\bibnamefont {{Y. Zhang, J. F.
  Douglas, B. D. Ermi, and E. J. Amis}}},\ }\href@noop {} {\bibfield  {journal}
  {\bibinfo  {journal} {{J. Chem. Phys.}}\ }\textbf {\bibinfo {volume} {114}},\
  \bibinfo {pages} {3299--3313} (\bibinfo {year} {2001})}\BibitemShut {NoStop}%
\bibitem [{\citenamefont {{M. Sedlak}}(2002)}]{2002_upturn_contr_sedlak}%
  \BibitemOpen
  \bibfield  {author} {\bibinfo {author} {\bibnamefont {{M. Sedlak}}},\
  }\href@noop {} {\bibfield  {journal} {\bibinfo  {journal} {{J. Chem. Phys.}}\
  }\textbf {\bibinfo {volume} {116}},\ \bibinfo {pages} {5256--5262} (\bibinfo
  {year} {2002})}\BibitemShut {NoStop}%
\bibitem [{\citenamefont {{K. Dusek, and W.
  Prins}}(1969)}]{1969_Clustering_Dusek}%
  \BibitemOpen
  \bibfield  {author} {\bibinfo {author} {\bibnamefont {{K. Dusek, and W.
  Prins}}},\ }\href@noop {} {\bibfield  {journal} {\bibinfo  {journal} {{Adv.
  Polym. Sci.}}\ }\textbf {\bibinfo {volume} {6}} (\bibinfo {year}
  {1969})}\BibitemShut {NoStop}%
\bibitem [{\citenamefont {{E. Geissler, A. Hecht, and R.
  Duplessix}}(1982)}]{1982_cluster_Gelssier}%
  \BibitemOpen
  \bibfield  {author} {\bibinfo {author} {\bibnamefont {{E. Geissler, A. Hecht,
  and R. Duplessix}}},\ }\href@noop {} {\bibfield  {journal} {\bibinfo
  {journal} {{J. Polym. Sci. Part B Polym. Phys.}}\ }\textbf {\bibinfo {volume}
  {20}} (\bibinfo {year} {1982})}\BibitemShut {NoStop}%
\bibitem [{\citenamefont {{E. Geissler, F. Horkay, and A.
  Hecht}}(1993)}]{1993_Geissler_Horkay_Cluster}%
  \BibitemOpen
  \bibfield  {author} {\bibinfo {author} {\bibnamefont {{E. Geissler, F.
  Horkay, and A. Hecht}}},\ }\href@noop {} {\bibfield  {journal} {\bibinfo
  {journal} {{Phys. Rev. Lett.}}\ }\textbf {\bibinfo {volume} {71}} (\bibinfo
  {year} {1993})}\BibitemShut {NoStop}%
\bibitem [{\citenamefont {{J. M. Guenet, M. Kein, and A.
  Menelle}}(1989)}]{1989_Clustering_Guenet}%
  \BibitemOpen
  \bibfield  {author} {\bibinfo {author} {\bibnamefont {{J. M. Guenet, M. Kein,
  and A. Menelle}}},\ }\href@noop {} {\bibfield  {journal} {\bibinfo  {journal}
  {{Macromolecules}}\ }\textbf {\bibinfo {volume} {22}} (\bibinfo {year}
  {1989})}\BibitemShut {NoStop}%
\bibitem [{\citenamefont {Ruzicka}\ \emph {et~al.}(2011)\citenamefont
  {Ruzicka}, \citenamefont {Zaccarelli}, \citenamefont {Zulian}, \citenamefont
  {Angelini}, \citenamefont {Sztucki}, \citenamefont {Moussa{\"\i}d},
  \citenamefont {Narayanan},\ and\ \citenamefont
  {Sciortino}}]{ruzicka2011observation}%
  \BibitemOpen
  \bibfield  {author} {\bibinfo {author} {\bibfnamefont {B.}~\bibnamefont
  {Ruzicka}}, \bibinfo {author} {\bibfnamefont {E.}~\bibnamefont {Zaccarelli}},
  \bibinfo {author} {\bibfnamefont {L.}~\bibnamefont {Zulian}}, \bibinfo
  {author} {\bibfnamefont {R.}~\bibnamefont {Angelini}}, \bibinfo {author}
  {\bibfnamefont {M.}~\bibnamefont {Sztucki}}, \bibinfo {author} {\bibfnamefont
  {A.}~\bibnamefont {Moussa{\"\i}d}}, \bibinfo {author} {\bibfnamefont
  {T.}~\bibnamefont {Narayanan}}, \ and\ \bibinfo {author} {\bibfnamefont
  {F.}~\bibnamefont {Sciortino}},\ }\href@noop {} {\bibfield  {journal}
  {\bibinfo  {journal} {Nat. Mater.}\ }\textbf {\bibinfo {volume} {10}},\
  \bibinfo {pages} {56--60} (\bibinfo {year} {2011})}\BibitemShut {NoStop}%
\bibitem [{\citenamefont {SantaLucia}(1998)}]{santalucia1998unified}%
  \BibitemOpen
  \bibfield  {author} {\bibinfo {author} {\bibfnamefont {J.}~\bibnamefont
  {SantaLucia}},\ }\href@noop {} {\bibfield  {journal} {\bibinfo  {journal}
  {Proc. Natl. Acad. Sci. USA}\ }\textbf {\bibinfo {volume} {95}},\ \bibinfo
  {pages} {1460--1465} (\bibinfo {year} {1998})}\BibitemShut {NoStop}%
\bibitem [{\citenamefont {Van~der Maarel}(2008)}]{van2008introduction}%
  \BibitemOpen
  \bibfield  {author} {\bibinfo {author} {\bibfnamefont {J.~R.}\ \bibnamefont
  {Van~der Maarel}},\ }\href@noop {} {\emph {\bibinfo {title} {Introduction to
  biopolymer physics}}}\ (\bibinfo  {publisher} {World Sci.},\ \bibinfo {year}
  {2008})\BibitemShut {NoStop}%
\bibitem [{\citenamefont {Sciortino}\ and\ \citenamefont
  {Zaccarelli}(2011)}]{sciortino2011reversible}%
  \BibitemOpen
  \bibfield  {author} {\bibinfo {author} {\bibfnamefont {F.}~\bibnamefont
  {Sciortino}}\ and\ \bibinfo {author} {\bibfnamefont {E.}~\bibnamefont
  {Zaccarelli}},\ }\href@noop {} {\bibfield  {journal} {\bibinfo  {journal}
  {Curr. Opin. Solid St. M.}\ }\textbf {\bibinfo {volume} {15}},\ \bibinfo
  {pages} {246--253} (\bibinfo {year} {2011})}\BibitemShut {NoStop}%
\bibitem [{\citenamefont {{L. Rovigatti, F. Bomboi, and F.
  Sciortino}}(2014)}]{2014_Accurate_phase_diagram_4valent_DNA-NS}%
  \BibitemOpen
  \bibfield  {author} {\bibinfo {author} {\bibnamefont {{L. Rovigatti, F.
  Bomboi, and F. Sciortino}}},\ }\href@noop {} {\bibfield  {journal} {\bibinfo
  {journal} {{J. Chem. Phys.}}\ }\textbf {\bibinfo {volume} {140}} (\bibinfo
  {year} {2014})}\BibitemShut {NoStop}%
\bibitem [{\citenamefont {{L. Z. Rogovina, V. G. Vasil'ev, and E. E.
  Braudo}}(2008)}]{2008_Definition_PolymerGels}%
  \BibitemOpen
  \bibfield  {author} {\bibinfo {author} {\bibnamefont {{L. Z. Rogovina, V. G.
  Vasil'ev, and E. E. Braudo}}},\ }\href@noop {} {\bibfield  {journal}
  {\bibinfo  {journal} {{Polym. Sci. Ser. C}}\ }\textbf {\bibinfo {volume}
  {50}},\ \bibinfo {pages} {85--92} (\bibinfo {year} {2008})}\BibitemShut
  {NoStop}%
\bibitem [{\citenamefont {{T. Rossow, and S.
  Seiffert}}(2014)}]{2014_Supramolecular_polymer_gels}%
  \BibitemOpen
  \bibfield  {author} {\bibinfo {author} {\bibnamefont {{T. Rossow, and S.
  Seiffert}}},\ }\href@noop {} {\bibfield  {journal} {\bibinfo  {journal}
  {{Polym. Chem.}}\ }\textbf {\bibinfo {volume} {5}},\ \bibinfo {pages}
  {3018--3029} (\bibinfo {year} {2014})}\BibitemShut {NoStop}%
\bibitem [{\citenamefont {{S. Chaterji, I. K. Kwon, and K.
  Park}}(2007)}]{2007_Smart_Polymeric_Gels}%
  \BibitemOpen
  \bibfield  {author} {\bibinfo {author} {\bibnamefont {{S. Chaterji, I. K.
  Kwon, and K. Park}}},\ }\href@noop {} {\bibfield  {journal} {\bibinfo
  {journal} {{Prog. Polym. Sci.}}\ }\textbf {\bibinfo {volume} {32}},\ \bibinfo
  {pages} {1083--1122} (\bibinfo {year} {2007})}\BibitemShut {NoStop}%
\bibitem [{\citenamefont {{T. Matsunaga, T. Sakai, Y. Akagi, U. il Chung, and
  M. Shibayama}}(2009)}]{2009_Shibayama_Matsunaga}%
  \BibitemOpen
  \bibfield  {author} {\bibinfo {author} {\bibnamefont {{T. Matsunaga, T.
  Sakai, Y. Akagi, U. il Chung, and M. Shibayama}}},\ }\href@noop {} {\bibfield
   {journal} {\bibinfo  {journal} {{Macromolecules}}\ }\textbf {\bibinfo
  {volume} {42}},\ \bibinfo {pages} {6245--6252} (\bibinfo {year}
  {2009})}\BibitemShut {NoStop}%
\bibitem [{\citenamefont {{L. H. Lee}}(1989)}]{1989_BOOK_Lee_NewTrends}%
  \BibitemOpen
  \bibfield  {author} {\bibinfo {author} {\bibnamefont {{L. H. Lee}}},\
  }\href@noop {} {\bibfield  {journal} {\bibinfo  {journal} {{Xerox Coorp.
  Plenum Press, 1st ed.}}\ ,\ \bibinfo {pages} {483--495}} (\bibinfo {year}
  {1989})}\BibitemShut {NoStop}%
\bibitem [{\citenamefont {{H. Benoit, D. Decker, C. Duplessix, C. Picot, and P.
  Rempp}}(1976)}]{1976_Benoit_PolymNetworks}%
  \BibitemOpen
  \bibfield  {author} {\bibinfo {author} {\bibnamefont {{H. Benoit, D. Decker,
  C. Duplessix, C. Picot, and P. Rempp}}},\ }\href@noop {} {\bibfield
  {journal} {\bibinfo  {journal} {{J. Polym. Sci. Part B Polym. Phys.}}\
  }\textbf {\bibinfo {volume} {14}},\ \bibinfo {pages} {2199--2128} (\bibinfo
  {year} {1976})}\BibitemShut {NoStop}%
\bibitem [{\citenamefont {{R. Ullman}}(1982)}]{1982_Ullman_Macromolecules}%
  \BibitemOpen
  \bibfield  {author} {\bibinfo {author} {\bibnamefont {{R. Ullman}}},\
  }\href@noop {} {\bibfield  {journal} {\bibinfo  {journal} {{Macromolecules}}\
  }\textbf {\bibinfo {volume} {15}},\ \bibinfo {pages} {1395--1402} (\bibinfo
  {year} {1982})}\BibitemShut {NoStop}%
\bibitem [{\citenamefont {{H. M. Tsay, and R.
  Ullman}}(1988)}]{1988_Tsay_Ullman}%
  \BibitemOpen
  \bibfield  {author} {\bibinfo {author} {\bibnamefont {{H. M. Tsay, and R.
  Ullman}}},\ }\href@noop {} {\bibfield  {journal} {\bibinfo  {journal}
  {{Macromolecules}}\ }\textbf {\bibinfo {volume} {21}},\ \bibinfo {pages}
  {2963} (\bibinfo {year} {1988})}\BibitemShut {NoStop}%
\bibitem [{\citenamefont {{A. M. Jonker, D. W. P. M. L{\"o}wik, and J. C. M.
  van Hest}}(2012)}]{2012_Peptide_protein_hydrogels}%
  \BibitemOpen
  \bibfield  {author} {\bibinfo {author} {\bibnamefont {{A. M. Jonker, D. W. P.
  M. L{\"o}wik, and J. C. M. van Hest}}},\ }\href@noop {} {\bibfield  {journal}
  {\bibinfo  {journal} {{Chem. Mater.}}\ }\textbf {\bibinfo {volume} {24}},\
  \bibinfo {pages} {759--773} (\bibinfo {year} {2012})}\BibitemShut {NoStop}%
\bibitem [{\citenamefont {{J. D. Ehrick, S. K. Deo, T. W. Browning, L. G.
  Bachas, M. J. Madou, and S.
  Daunert}}(2005)}]{2005_Genetically_engineered_protein_hydrogels}%
  \BibitemOpen
  \bibfield  {author} {\bibinfo {author} {\bibnamefont {{J. D. Ehrick, S. K.
  Deo, T. W. Browning, L. G. Bachas, M. J. Madou, and S. Daunert}}},\
  }\href@noop {} {\bibfield  {journal} {\bibinfo  {journal} {{Nat. Mater.}}\
  }\textbf {\bibinfo {volume} {4}},\ \bibinfo {pages} {298--302} (\bibinfo
  {year} {2005})}\BibitemShut {NoStop}%
\bibitem [{\citenamefont {{S. Tang, M. Wang, and B. D.
  Olsen}}(2015)}]{2015_Associative_Protein_Hydrogels}%
  \BibitemOpen
  \bibfield  {author} {\bibinfo {author} {\bibnamefont {{S. Tang, M. Wang, and
  B. D. Olsen}}},\ }\href@noop {} {\bibfield  {journal} {\bibinfo  {journal}
  {{J. Am. Chem. Soc.}}\ }\textbf {\bibinfo {volume} {137}},\ \bibinfo {pages}
  {3946--3957} (\bibinfo {year} {2015})}\BibitemShut {NoStop}%
\bibitem [{\citenamefont {{P. Li, Siddaramaiah, N. H. Kim, S. B. Heo, and J. H.
  Lee}}(2008)}]{2008_PAAMlaponite_Clay_Nanocomposite}%
  \BibitemOpen
  \bibfield  {author} {\bibinfo {author} {\bibnamefont {{P. Li, Siddaramaiah,
  N. H. Kim, S. B. Heo, and J. H. Lee}}},\ }\href@noop {} {\bibfield  {journal}
  {\bibinfo  {journal} {{Composites Part B}}\ }\textbf {\bibinfo {volume}
  {39}},\ \bibinfo {pages} {756--763} (\bibinfo {year} {2008})}\BibitemShut
  {NoStop}%
\bibitem [{\citenamefont {{C. W. Chang, A. Van Spreeuwel, C. Zhang, and S.
  Varghese}}(2010)}]{2010_PEG_Clay_nanocomposite_hydrogel}%
  \BibitemOpen
  \bibfield  {author} {\bibinfo {author} {\bibnamefont {{C. W. Chang, A. Van
  Spreeuwel, C. Zhang, and S. Varghese}}},\ }\href@noop {} {\bibfield
  {journal} {\bibinfo  {journal} {{Soft Matter}}\ }\textbf {\bibinfo {volume}
  {6}},\ \bibinfo {pages} {5157--5164} (\bibinfo {year} {2010})}\BibitemShut
  {NoStop}%
\bibitem [{\citenamefont {{L. Z. Zhao, C. H. Zhou, J. Wang, D. S. Tong, W. H.
  Yu, and H. Wang}}(2015)}]{2015_Clay_Mineral_Nanocomposite_hydrogels}%
  \BibitemOpen
  \bibfield  {author} {\bibinfo {author} {\bibnamefont {{L. Z. Zhao, C. H.
  Zhou, J. Wang, D. S. Tong, W. H. Yu, and H. Wang}}},\ }\href@noop {}
  {\bibfield  {journal} {\bibinfo  {journal} {{Soft Matter}}\ }\textbf
  {\bibinfo {volume} {11}},\ \bibinfo {pages} {9229--9246} (\bibinfo {year}
  {2015})}\BibitemShut {NoStop}%
\bibitem [{\citenamefont {{M. Shibayama}}(2011)}]{2011_Shibayama}%
  \BibitemOpen
  \bibfield  {author} {\bibinfo {author} {\bibnamefont {{M. Shibayama}}},\
  }\href@noop {} {\bibfield  {journal} {\bibinfo  {journal} {{Polym. J.}}\
  }\textbf {\bibinfo {volume} {43}} (\bibinfo {year} {2011})}\BibitemShut
  {NoStop}%
\end{thebibliography}%

\end{document}